\documentclass[twocolumn]{aastex701}

\usepackage{amsmath}
\usepackage{url}
\usepackage{xspace}
\let\oldAA\AA
\renewcommand{\AA}{\text{\normalfont\oldAA}\xspace}
\newcommand{\eg}{\textit{e.g.}}

\usepackage{xcolor}
\newcommand{\target}{\object{UDS-27023}\xspace}

\shorttitle{Steep-Extinction QSO}
\shortauthors{M. Li et al.}
\graphicspath{{./}{figures/}}

\begin{document}

\title{
A Steep-Extinction Quasi-stellar Object at z=4.6: JWST Evidence for Abundant Small Dust Grains
}

\author[0000-0001-6251-649X]{Mingyu Li}
\affiliation{Department of Astronomy, Tsinghua University, Beijing 100084, China}
\affiliation{Kavli Institute for Cosmology, University of Cambridge, Madingley Road, Cambridge CB3 0HA, UK}
\affiliation{Cavendish Laboratory, University of Cambridge, 19 JJ Thomson Avenue, Cambridge CB3 0HE, UK}
\email[show]{lmytime@hotmail.com}

\author[0000-0001-8467-6478]{Zheng Cai}
\affiliation{Department of Astronomy, Tsinghua University, Beijing 100084, China}
\email{zcai@mail.tsinghua.edu.cn}

\author[0000-0002-4985-3819]{Roberto Maiolino}
\affiliation{Kavli Institute for Cosmology, University of Cambridge, Madingley Road, Cambridge CB3 0HA, UK}
\affiliation{Cavendish Laboratory, University of Cambridge, 19 JJ Thomson Avenue, Cambridge CB3 0HE, UK}
\affiliation{Department of Physics and Astronomy, University College London, Gower Street, London WC1E 6BT, UK}
\email{rm665@cam.ac.uk}

\author[0000-0002-4622-6617]{Fengwu Sun}
\affiliation{Center for Astrophysics $|$ Harvard \& Smithsonian, 60 Garden St., Cambridge, MA 02138, USA}
\affiliation{Department of Astronomy, Westlake University, Hangzhou 310030, People’s Republic of China}
\email{sunfengwu@westlake.edu.cn}

\author[0000-0002-1660-9502]{Xihan Ji}
\affiliation{Kavli Institute for Cosmology, University of Cambridge, Madingley Road, Cambridge CB3 0HA, UK}
\affiliation{Cavendish Laboratory, University of Cambridge, 19 JJ Thomson Avenue, Cambridge CB3 0HE, UK}
\email{xj274@cam.ac.uk}

\author[0009-0009-8105-4564]{Qiao Duan}
\affiliation{Kavli Institute for Cosmology, University of Cambridge, Madingley Road, Cambridge CB3 0HA, UK}
\affiliation{Cavendish Laboratory, University of Cambridge, 19 JJ Thomson Avenue, Cambridge CB3 0HE, UK}
\email{qd231@cam.ac.uk}

\author[0000-0003-2983-815X]{Bjorn\,H.\,C. Emonts}
\affiliation{National Radio Astronomy Observatory, 520 Edgemont Road, Charlottesville, VA 22903, USA}
\email{bemonts@nrao.edu}

\author[0000-0003-3310-0131]{Xiaohui Fan}
\affiliation{Steward Observatory, University of Arizona, 933 N Cherry Avenue, Tucson, AZ 85721, USA}
\email{xiaohuidominicfan@gmail.com}

\author[0009-0003-7423-8660]{Ignas Juodžbalis}
\affiliation{Kavli Institute for Cosmology, University of Cambridge, Madingley Road, Cambridge CB3 0HA, UK}
\affiliation{Cavendish Laboratory, University of Cambridge, 19 JJ Thomson Avenue, Cambridge CB3 0HE, UK}
\email{ĳ284@cam.ac.uk}

\author[0000-0001-6052-4234]{Xiaojing Lin}
\affiliation{Department of Astronomy, Tsinghua University, Beijing 100084, China}
\email{linxj21@mails.tsinghua.edu.cn}

\author[0009-0006-4990-7529]{Yixiao Liu}
\affiliation{Chinese Academy of Sciences South America Center for Astronomy (CASSACA), National Astronomical Observatories (NAOC), 20A Datun Road, Beijing 100012, China}
\affiliation{School of Astronomy and Space Science, University of Chinese Academy of Sciences, Beijing 101408, China}
\affiliation{Kavli Institute for Cosmology, University of Cambridge, Madingley Road, Cambridge CB3 0HA, UK}
\affiliation{Cavendish Laboratory, University of Cambridge, 19 JJ Thomson Avenue, Cambridge CB3 0HE, UK}
\email{liuyixiao@nao.cas.cn}

\author[0000-0002-8224-4505]{Sandro Tacchella}
\affiliation{Kavli Institute for Cosmology, University of Cambridge, Madingley Road, Cambridge CB3 0HA, UK}
\affiliation{Cavendish Laboratory, University of Cambridge, 19 JJ Thomson Avenue, Cambridge CB3 0HE, UK}
\email{st578@cam.ac.uk}


\begin{abstract}
The rapid accumulation of massive dust reservoirs in the early Universe remains a major challenge in astrophysics. While core-collapse supernovae can inject large dust grains ($a \gtrsim 0.1\,\mu{\rm m}$) on short timescales, explaining the total dust budgets in the early Universe likely requires efficient grain growth in the interstellar medium (ISM). Such growth depends critically on an abundant population of small grains, which maximize the surface area available for accretion and may be generated by rapid dust-processing or dust-formation channels. Here, we report the discovery of a QSO, UDS-27023, at $z=4.556\pm0.003$, identified using JWST/NIRSpec spectroscopy. By quantitatively comparing the spectra to QSO composite templates, we find that UDS-27023 displays an exceptionally steep far-UV extinction curve ($A_{1500}/A_V \approx 8$) but notably lacks the 2175 \AA bump {($A_\mathrm{bump}/A_V<0.34$ at $3\sigma$)}, indicating a dominance of small silicate dust grains. We interpret this phenomenology as evidence for active small-grain production and processing in the QSO environment. Mechanical shattering of pre-existing large grains by QSO-driven shocks and outflows provides one natural pathway, while in situ condensation of silicate grains inside dense QSO-driven winds may offer an additional route. Such a population of steep-extinction QSOs (SEQs) may therefore reveal a short-lived phase in which luminous active galactic nuclei generate, process, and redistribute small grains, potentially facilitating rapid ISM grain growth and enriching the circumgalactic medium.
\end{abstract}



\section{Introduction} \label{sec:intro}

Dust is a fundamental constituent of galaxies across cosmic time, regulating the thermodynamic state of the gas, catalyzing the formation of molecular clouds, and obscuring star formation tracers \citep{Draine2003ARA&A..41..241D, Li2003ssac.proc...37L}.
Understanding the origin and evolution of dust in the first billion years of the Universe remains one of the most contentious frontiers in astrophysics \citep{Schneider2024A&ARv..32....2S}.
One of the primary diagnostics is the dust grain size distribution, which fundamentally dictates both the total dust mass budget and the slope of the extinction curve \citep[\eg,][]{Hirashita2012MNRAS.422.1263H, Asano2013MNRAS.432..637A}.
Specifically, an abundance of small grains leads to a steep rise in extinction at ultraviolet (UV) wavelengths, whereas a distribution dominated by large grains results in a flatter, ``grey'' extinction curve.

The initial injection of dust into the early interstellar medium (ISM) is believed to be dominated by core-collapse supernovae (CCSNe), which typically produce large grains ($a > 0.1\,\mu{\rm m}$) on short timescales \citep[$< 10$~Myr; \eg,][]{Todini2001MNRAS.325..726T, Maiolino2004Natur.431..533M, Gall2011A&A...528A..13G}.
This large-grain dominance is supported by recent JWST spectroscopic studies, which show that dust attenuation curves at higher redshift are generally flatter than those at lower redshift \citep{Markov2025NatAs...9..458M, Shivaei2025arXiv250901795S}.
However, this creates a tension with the total dust mass observed in the epoch of reionization (EoR).
Over the past decade, ALMA observations have revealed surprisingly mature dust reservoirs ($\gtrsim10^7 M_{\odot}$) in galaxies as early as $z \sim 7-8$ \citep[\eg,][]{Riechers2013Natur.496..329R, Watson2015Natur.519..327W, Tamura2019ApJ...874...27T, Inami2022MNRAS.515.3126I, Algera2026MNRAS.545f1897A}, initiating a long-standing debate known as the ``dust budget crisis'' \citep[\eg,][]{Valiante2011MNRAS.416.1916V, Rowlands2014MNRAS.441.1040R}.
While some theoretical models propose that maximal supernova efficiency, combined with minimal destruction, could marginally account for these massive dust reservoirs \citep[\eg,][]{Lesniewska2019A&A...624L..13L}, the consensus increasingly points to the necessity of rapid ISM grain growth to explain these masses.

To facilitate rapid grain growth, the ISM requires a large surface area-to-mass ratio, which can be achieved through an abundance of small grains \citep{Kuo2012MNRAS.424L..34K, Asano2013MNRAS.432..637A}.
The presence of such small grains in the EoR has recently been suggested by the detection of the 2175~\AA bump, a feature attributed to small carbonaceous dust grains, such as polycyclic aromatic hydrocarbons (PAHs) or graphite \citep{Witstok2023Natur.621..267W, Markov2025NatAs...9..458M, Fisher2025MNRAS.539..109F}.
However, identifying the origin of these grains remains complex.
In the early Universe, where the contribution from asymptotic giant branch (AGB) stars is limited by evolutionary timescales \citep[\eg,][]{Valiante2009MNRAS.397.1661V}, small grains may be produced through several rapid channels, including mechanical shattering of large supernova-produced grains via grain--grain collisions in turbulent environments \citep[\eg,][]{Hirashita2010MNRAS.407L..49H} and, in luminous AGN, direct dust formation within dense QSO-driven winds \citep[\eg,][]{Elvis2002ApJ...567L.107E}.
These channels have two important consequences: they increase the grain surface area available for accretion, thereby catalyzing dust-mass assembly within galaxies, and they generate small-grain populations that can be entrained in outflows and expelled into the halo.
This latter effect has recently been inferred by JWST observations of dusty star-forming galaxies (DSFGs) at $z\sim3.5$ \citep{Sun2026arXiv260115961S}, which reveal large reservoirs of small silicate grains in the circumgalactic medium (CGM).

However, linking the massive dust reservoirs in DSFGs to these enriched halos requires identifying the active phase of small-grain production and processing.
In this work, we present observational evidence for such a phase in the QSO UDS-27023 at $z=4.6$, identified via JWST/NIRSpec spectroscopy.
Unlike typical high-redshift sources, this object exhibits an extremely steep extinction curve without a 2175 \AA\ bump, signaling a violent environment dominated by small silicate dust grains.
We argue that such a category of steep-extinction QSO (SEQ) may mark a transitional phase in which AGN feedback both processes pre-existing grains and possibly forms new grains within QSO-driven winds, while redistributing the resulting small-grain population into the ISM and CGM.
In Section~\ref{sec:obs}, we describe observational data and analysis.
We present the result of a steep dust extinction curve in Section~\ref{sec:result} and explore the interpretation in Section~\ref{sec:discussion}.
Finally, we summarize our major results and conclusions in Section~\ref{sec:summary}.
Throughout this work, magnitudes are given in the AB system \citep{Oke1983ApJ...266..713O}.
We adopt a flat $\mathrm{\Lambda CDM}$ cosmology with $\Omega_{\mathrm{m}}=0.3$ and $H_{0}=70 \mathrm{~km}\mathrm{~s}^{-1} \mathrm{~Mpc}^{-1}$.
In this cosmology, 1\arcsec\ corresponds to 6.57 kpc of physical length at $z=4.556$.
The pivot wavelengths of the V- and B-bands are 5500~\AA and 4400~\AA, respectively.

\begin{figure*}[t]
    \centering
    \includegraphics[width=\linewidth]{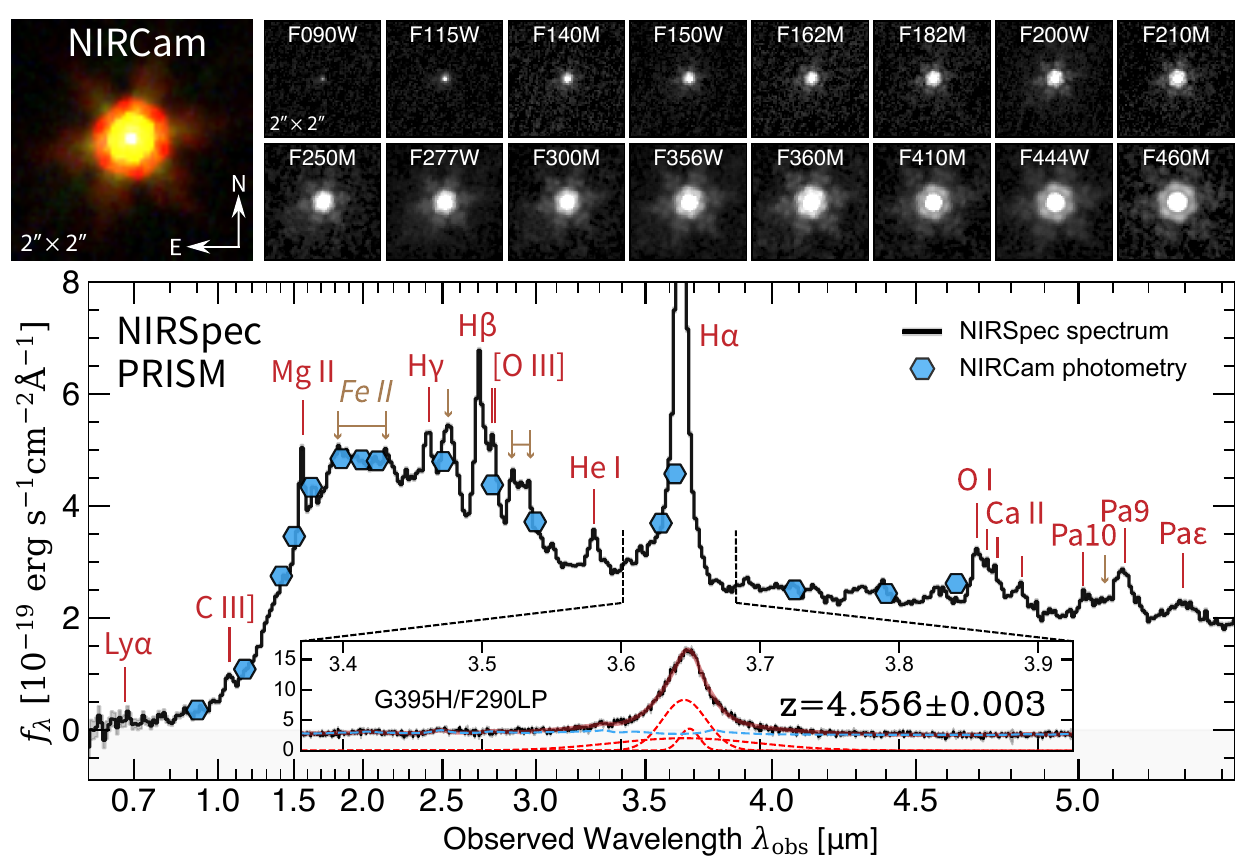}
    \caption{
    JWST imaging and spectroscopy of \target.
    \textit{Top:} $2\arcsec\times2\arcsec$ JWST/NIRCam cutouts in the available filters and an RGB composite (left), showing that the source is compact and dominated by a point source in all bands.
    \textit{Bottom:} JWST/NIRSpec PRISM spectrum (black) overlaid with the NIRCam photometry (blue hexagons), illustrating the bright continuum and strong broad emission lines characteristic of a Type~I QSO. The spectrum shows a strongly suppressed rest-frame UV continuum relative to the rest-frame optical emission, suggesting an exceptionally steep extinction curve.
    \textit{Inset:} Zoom-in on the G395H/F290LP spectrum around H$\alpha$, together with a multi-component fit (solid red; dashed components). The spectroscopic redshift of \target is measured to be $z=4.556\pm0.003$.
}
    \label{fig:imgspec}
\end{figure*}

\begin{figure*}[t]
\centering
\includegraphics[width=\linewidth]{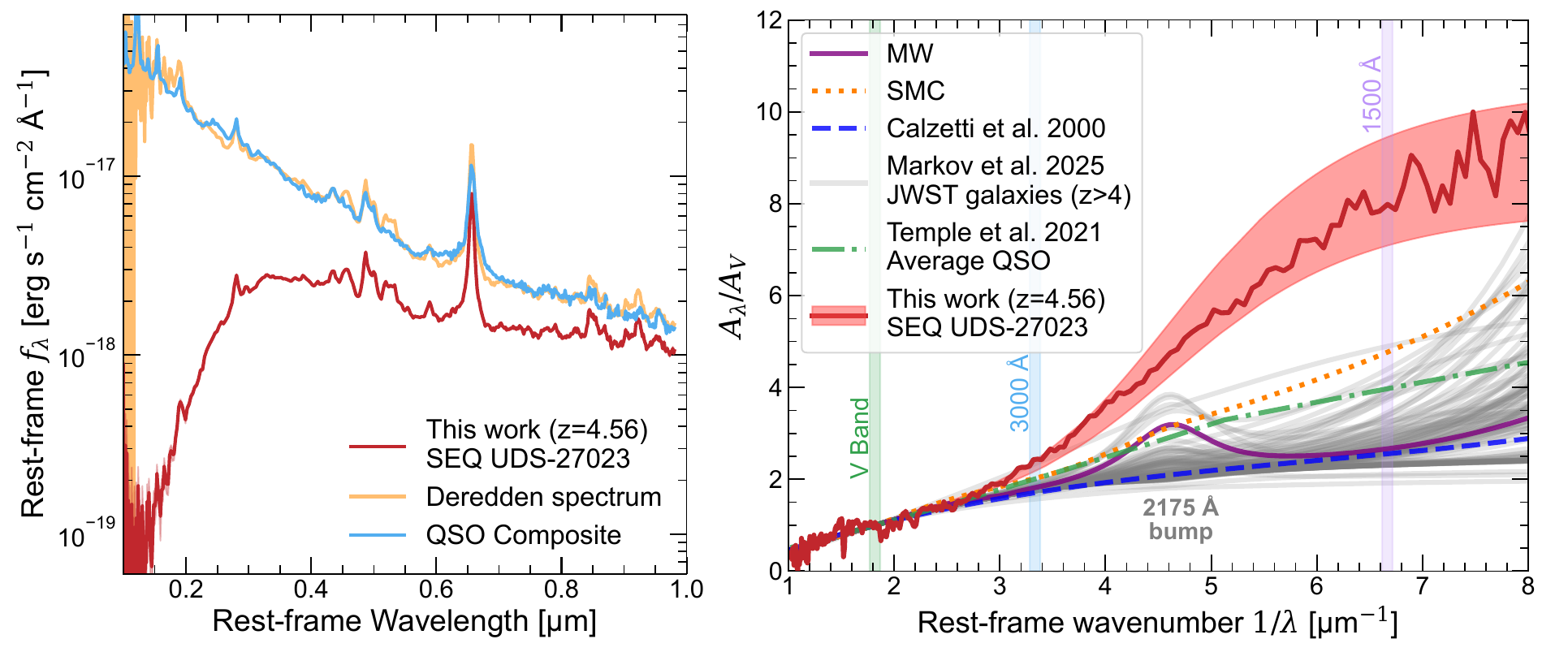}
\caption{
Rest-frame SED and inferred line-of-sight extinction curve of \target.
\textit{Left:} Rest-frame UV--optical--NIR spectrum of \target (red) compared to an intrinsic blue-QSO composite (blue). The orange curve shows the spectrum corrected using the best-fit extinction curve, illustrating that the dereddened continuum approaches the intrinsic QSO template over the wavelength range.
\textit{Right:} Extinction curve reconstructed for \target\ shown as $A_\lambda/A_\mathrm{V}$ versus wavenumber ($1/\lambda$; thick red curve), with the shaded red band indicating the corresponding best-fit parametric extinction model with uncertainty. For comparison, we plot the Milky Way and SMC extinction curves, the \citet{Calzetti2000ApJ...533..682C} starburst attenuation law, the average QSO extinction curve \citep{Temple2021MNRAS.508..737T}, and the set of attenuation curves measured for JWST\ galaxies at $z>4$ by \citealt{Markov2025NatAs...9..458M} (thin grey curves). Vertical shaded bands mark the $V$ band and rest-frame 3000~\AA\ and 1500~\AA; the location of the 2175~\AA\ bump is indicated for reference. The SEQ extinction curve rises substantially more steeply into the UV than any of these commonly used laws.
}
\label{fig:dustlaw}
\end{figure*}

\begin{figure*}[t]
\centering
\includegraphics[width=\linewidth]{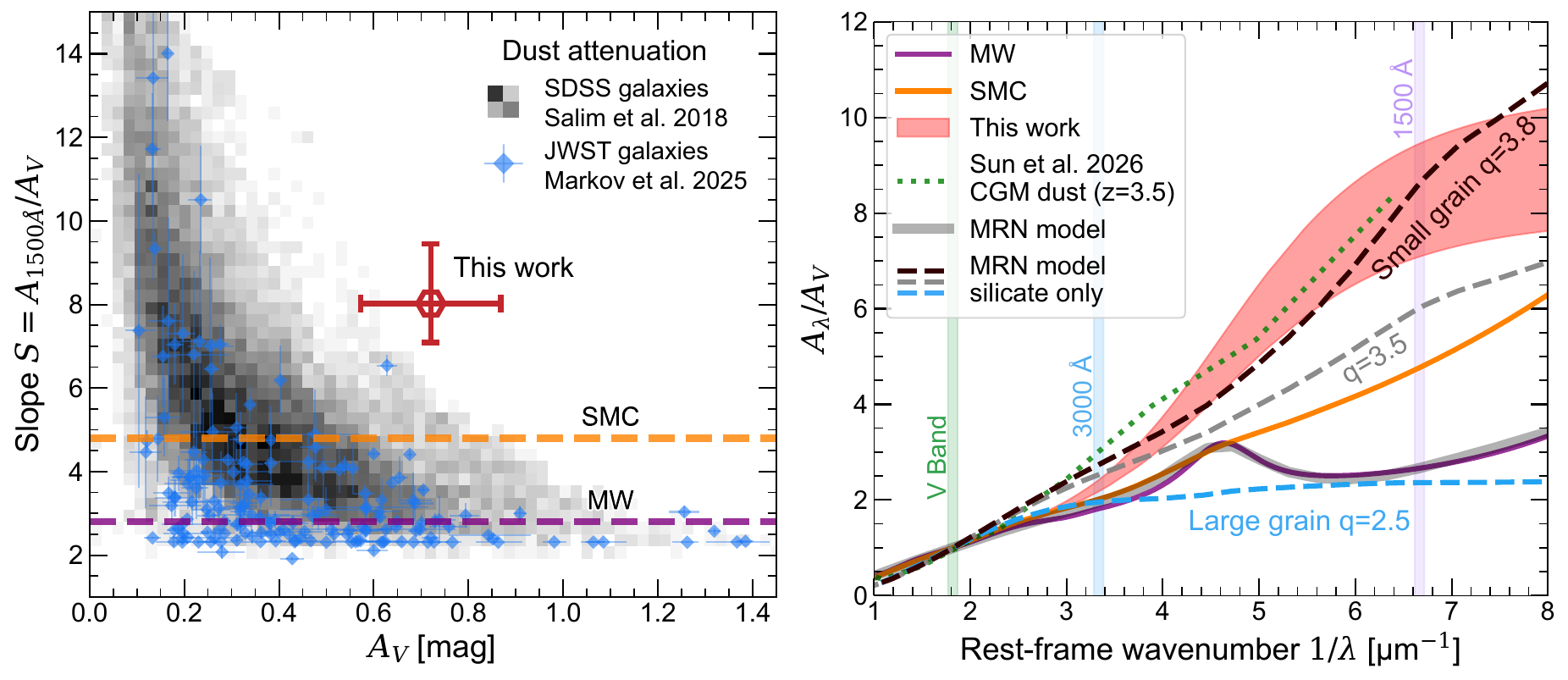}
\caption{
Physical interpretation and context for the steep extinction toward \target.
\textit{Left:} Location of SEQ (red symbol) in the plane of dust-law steepness, $S\equiv A_{1500}/A_\mathrm{V}$, versus $A_\mathrm{V}$.
The grey-scale background shows the distribution of dust attenuation slopes inferred for SDSS galaxies from the GSWLC survey \citep{Salim2016ApJS..227....2S, Salim2018ApJ...859...11S}, and the blue points show the $z\approx2$--10 JWST galaxy attenuation constraints from \citet{Markov2025NatAs...9..458M}.
Horizontal dashed lines indicate the values expected for standard Milky Way and SMC extinction curves, illustrating that SEQ lies at significantly larger $S$ than both typical low- and high-redshift galaxy extinction and attenuation laws at moderate $A_\mathrm{V}$.
{We note that this work (red point) provides a line-of-sight \textit{extinction} measurement toward an unresolved point source, whereas the grey and blue distributions are \textit{attenuation} constraints for spatially extended galaxies; the two are not strictly commensurable, and the comparison should be read in that light (Section~\ref{sec:discussion}).}
\textit{Right:} Comparison of the SEQ extinction curve (red band; normalized at V-band) to common empirical laws (MW and SMC) and to illustrative MRN grain-size models \citep{Mathis1977ApJ...217..425M} computed from optical properties of silicate and graphite \citep{Draine1984ApJ...285...89D, Weingartner2001ApJ...548..296W, Laor1993ApJ...402..441L}.
The grey curves show an MRN-like size distribution ($dn/da\propto a^{-3.5}$) with a standard silicate+graphite mixture (solid) and silicate-only dust (dashed).
Colored curves demonstrate how varying the size-distribution slope $q$ (defined by $dn/da\propto a^{-q}$) changes the UV rise: small-grain--weighted distributions (larger $q$) produce substantially steeper far-UV extinction than large-grain/gray cases (smaller $q$).
The green dotted line denotes a CGM dust extinction curve with a steep slope \citep{Sun2026arXiv260115961S} at $z=3.5$.
Vertical bands mark rest-frame 3000~\AA\ and 1500~\AA\ and the $V$ band.
Together, these panels show that SEQ probes an extreme, line-of-sight extinction law, qualitatively consistent with a dust population strongly weighted toward small grains.
}
\label{fig:dustmodel}
\end{figure*}

\section{Data and Analysis} \label{sec:obs}
The target \target is located in the UKIDSS Ultra Deep Survey (UDS) field \citep{Lawrence2007MNRAS.379.1599L}, which is the core region of the Subaru-XMM Deep Field \citep[SXDS;][]{Furusawa2008ApJS..176....1F}.
The equatorial coordinates (J2000) of this target are $\rm R.A.=34.33730~deg$, $\rm Decl.=-5.14365~deg$.
The source ID 27023 is from the CANDELS multiwavelength catalog \citep{Galametz2013ApJS..206...10G}.
Before JWST observation, \target at $z_\mathrm{spec}=4.556\pm0.003$ was identified as a quiescent galaxy hosting an IR-AGN at $z_\mathrm{phot}=2-3$ \citep{Bruce2012MNRAS.427.1666B, Cowley2016MNRAS.457..629C, Patil2019ApJ...871..109P, Carnall2020MNRAS.496..695C, Kodra2023ApJ...942...36K}.
Even when using the \texttt{Prospector} SED modeling framework \citep{Johnson2021ApJS..254...22J} on the JWST photometry, with a non-parametric star formation history and the star-forming main sequence (SFMS) prior \citep{Duan2026arXiv260521599D}, \target is still classified as a quiescent or post-starburst galaxy at $z_\mathrm{phot}=2-3$.
We further discuss this underestimated photometric redshift in Appendix~\ref{app:photz}.

\subsection{JWST imaging and spectroscopy}

We use the JWST/NIRCam imaging observed by two programs: PRIMER (GO\#1837; PI: J. Dunlop) and MINERVA (GO\#7814; PIs: A. Muzzin, D. Marchesini, and K. Suess).
We download the calibrated single exposures (\texttt{\_cal.fits}) from the MAST data archive and apply customized data-reduction steps, including WISP removal, 1/f noise removal, sky background subtraction, and astrometric correction. 
The final mosaic images are drizzled to a pixel size of 0\farcs03.
The $2\arcsec\times2\arcsec$ cutout images in each filter band and an RGB image are shown in the top panels of Fig.~\ref{fig:imgspec}.
We find that in all bands, the source is point-source-dominated.
We perform aperture photometry within EE80-radius apertures (see Table~3 in \href{https://jwst-docs.stsci.edu/jwst-near-infrared-camera/nircam-performance/nircam-point-spread-functions}{JWST-DOCS}), and then an 80\% aperture correction is applied to obtain the total fluxes.

We utilize the JWST/NIRSpec MSA spectroscopic data in PRISM and G395H/F290LP modes as part of the GTO\#1215 project \citep[PI: N. Luetzgendorf;][]{Maseda2024A&A...689A..73M}.
The reduced spectra are retrieved from the Dawn JWST Archive (DJA)\footnote{https://dawn-cph.github.io/dja}, which are reduced using the \texttt{MSAEXP} pipeline\footnote{https://github.com/gbrammer/msaexp} \citep{msaexp}, introduced by \citealt{Heintz2024Sci...384..890H, deGraaff2025A&A...697A.189D}.
We perform a slit-loss correction by matching the NIRSpec spectrum to the NIRCam photometry, scaling with a best-fit quadratic polynomial function.
The resulting NIRSpec spectra and NIRCam photometry are shown in the bottom panel of Fig.~\ref{fig:imgspec} and are publicly released on GitHub\footnote{\url{https://github.com/lmytime/JWST_UDS-27023}}.

JWST observations show that \target is a reddened Type~I QSO at $z\simeq4.56$ (Fig.~\ref{fig:imgspec}).
The NIRCam cutouts show that an unresolved point source dominates the emission in all available filters, and the NIRSpec spectra exhibit prominent Fe\,{\sc ii} emission and a series of broad permitted lines, characteristic of a high-$z$ Type~I QSO.
Compared with the normalized QSO composite spectrum \citep{Selsing2016A&A...585A..87S}, we find a strongly suppressed UV continuum blueward of rest-frame wavelength $\lambda_{\rm rest}\sim3000$~\AA\ relative to the intrinsic composite (Fig.~\ref{fig:dustlaw} left).
The reddening has two regimes: moderate in the rest-optical but extreme in the UV, suggesting that the dust extinction toward the QSO continuum rises much more steeply toward shorter wavelengths than canonical extinction laws.

To measure the basic QSO properties of \target, we model the H$\alpha$ complex observed in G395H grating (Fig.~\ref{fig:imgspec} inset).
We fit the spectrum with a linear continuum, a QSO Fe\,{\sc ii} template \citep{Veron-Cetty2004A&A...417..515V}, and multiple Gaussian components of H$\alpha$ emission.
All components are fit simultaneously.
The best-fitting model requires three broad H$\alpha$ components, yielding a total flux of $F_{\rm H\alpha}=(6.6\pm0.2)\times10^{-16}~\mathrm{erg~s^{-1}~cm^{-2}}$ and an effective line width of ${\rm FWHM}_{\rm H\alpha}=2803\pm \mathrm{49}~\mathrm{km~s^{-1}}$.
Notably, the spectrum is well reproduced without invoking a narrow H$\alpha$ component ($\lesssim500~\mathrm{km~s^{-1}}$) or forbidden lines from narrow line region such as [N~{\sc ii}], [S~{\sc ii}], and [O~{\sc i}]; adding these components does not statistically improve the fit, indicating that their contribution is negligible in this source.
In addition to the non-detection or negligible contribution of [N~{\sc ii}], [S~{\sc ii}] and [O~{\sc i}], we note that no high-ionization lines are detected in the NIRSpec spectra.
The spectroscopic redshift of \target is derived to be $z=4.556\pm0.003$, with uncertainty accounting for both fitting errors and multi-component velocity offsets.
Based on the assumed cosmology, the H$\alpha$ line luminosity is measured to be $L_\mathrm{H\alpha} = (1.38\pm0.04)\times10^{44}~\mathrm{erg\,s^{-1}}$.
We also measure the flux density at rest-frame 5100~\AA from the prism spectrum to be $f_{5100\AA} = (2.13\pm0.11)\times10^{-18}~\mathrm{erg\,s^{-1}\,cm^{-2}\,\AA^{-1}}$, which converts to a luminosity of $L_{5100\AA}=(2.28\pm0.11)\times10^{45}~\mathrm{erg\,s^{-1}}$.

Using the broad H$\alpha$ virial estimator based on ${\rm FWHM}_{\rm H\alpha}$ and $L_{\rm H\alpha}$ \citep{Greene2005ApJ...630..122G}, we infer a black-hole mass of $M_{\rm BH}=(2.5\pm0.5)\times10^{8}\,M_\odot$.
Assuming a typical bolometric correction of $\mathrm{BC}_\mathrm{5100}=6.43 \pm 1.84$ for Type I QSOs \citep{Chen2025ApJ...988..204C}, we estimate the QSO bolometric luminosity to be $L_\mathrm{bol}=(1.46\pm0.43)\times10^{46}~\mathrm{erg\,s^{-1}}$, which gives an Eddington ratio of $\lambda = L_\mathrm{bol}/L_\mathrm{edd}=0.5\pm0.2$.

\subsection{ALMA imaging}
We use ALMA data observed as part of two ALMA programs: 2012.1.00326.S (PI: Ikarashi) and 2015.1.01074.S (PI: Inami).
The ALMA spectral windows do not cover any significant emission lines, and the integrated continuum fluxes are measured to be $1.39\pm0.18$~mJy at 1.13~mm \citep[ALMA source name SXDF1100.027;][]{Ikarashi2015ApJ...810..133I, Patil2019ApJ...871..109P} and $2.84\pm0.53$~mJy at 0.87~mm (ALMA source name UDSb\_55).
The continuum emission is marginally resolved at both wavelengths. Image-plane morphology fitting yields source sizes of FWHM major and minor axes $0.46\pm0.05$ arcsec and $0.32\pm0.03$
arcsec (P.A.=$82.6^\circ\pm9.1^\circ$) at 1.13~mm, and $0.25\pm0.03$ arcsec and $0.19\pm0.02$ arcsec (P.A.=$-43.6^\circ\pm17.8^\circ$) at 0.87~mm.



We estimate the cold dust mass ($M_\mathrm{d}$) of the target galaxy using the detected dust continuum emission at 1.13~mm and 0.87~mm.
Since the available bands sample only the long-wavelength side of the thermal dust emission and leave the SED peak unconstrained, we cannot independently fit for the dust temperature ($T_\mathrm{d}$) or spectral index ($\beta$).
Instead, we derive a range of consistent dust masses by adopting a grid of canonical parameters typical for high-redshift star-forming galaxies with a modified blackbody spectrum ($\nu^\beta B_\nu$, where $B_\nu$ is a blackbody law).
We adopt a typical dust mass absorption coefficient of $\kappa_\mathrm{850\mu m}=0.077~\rm m^2\,kg^{-1}$ \citep{Draine1984ApJ...285...89D}, scaling it to the rest-frame wavelength of the observations as $\kappa_\nu\propto\nu^\beta$.
To account for the uncertainty in the intrinsic dust properties, we compute $M_\mathrm{d}$ over a parameter grid of $T_\mathrm{d}\in[25,55]$~K and $\beta\in[1.5,2.0]$, which encompasses the standard range of values observed in $z>4$ galaxies \citep[\eg,][]{Scoville2016ApJ...820...83S, Bethermin2020A&A...643A...2B}.
The resulting dust mass is $M_\mathrm{d}=(4.1\pm3.5)\times10^8~M_\odot$, with the uncertainty derived from the weighted average of the estimates across this grid, accounting for both photometric errors and systematic uncertainties in the assumed parameters.
This indicates a host galaxy with a rich interstellar dust environment.

\subsection{Other multi-wavelength data}
The Chandra X-UDS survey identifies a bright X-ray counterpart of \target (ID XUDS-270 in \citealt{Kocevski2018ApJS..236...48K}).
The observed X-ray flux is $\log (F_\mathrm{0.5-8\,keV}/\mathrm{erg\,s^{-1}\,cm^{-2}}) = -14.54$.
Assuming a power law index of $\Gamma=1.70$ and a column density of $\log N_\mathrm{H} / \mathrm{cm^{-2}} = 20.00$, the X-ray luminosity of \target is $\log(L_\mathrm{2-10\,keV}/\mathrm{erg\,s^{-1}})=44.42$ at $z=4.556$, consistent with the normal QSO distribution at this luminosity \citep{Duras2020A&A...636A..73D}.
This gives an X-ray bolometric correction factor of $k_\mathrm{X} = L_\mathrm{bol}/L_\mathrm{2-10\,keV}\approx55$.
In the radio regime, the source remains undetected down to an RMS of 1.6 $\mu$Jy/beam using VLA at 1--2~GHz \citep{Heywood2020MNRAS.496.3469H}.
Other multi-wavelength observations reveal bright rest-frame mid- to far-IR emission (JWST/MIRI, Spitzer/MIPS, Herschel-PACS/SPIRE).
A comprehensive SED decomposition incorporating these data is beyond the scope of this work.



\section{Steep dust extinction curve} \label{sec:result}

The visual comparison with the QSO composite template spectrum reveals a notable suppression of UV fluxes (left panel of Fig.~\ref{fig:dustlaw}), which strongly suggests a steep dust extinction along the line of sight of \target. 
Quantitatively, we model and characterize the dust extinction curve toward \target using the parametric formalism (the Drude approach) introduced by \citet{Li2008ApJ...685.1046L}.
The extinction law, normalized by $A_\mathrm{V}$, is expressed as a summation of Drude profiles:
\begin{equation}
\begin{aligned}
A_\lambda / A_\mathrm{V} & =\frac{c_1}{(\lambda / 0.08)^{c_2}+(0.08 / \lambda)^{c_2}+c_3} \\
& +\frac{233\left[1-c_1 /\left(6.88^{c_2}+0.145^{c_2}+c_3\right)-c_4 / 4.60\right]}{(\lambda / 0.046)^2+(0.046 / \lambda)^2+90} \\
& +\frac{c_4}{(\lambda / 0.2175)^2+(0.2175 / \lambda)^2-1.95},
\end{aligned}
\label{equ:1}
\end{equation}
where $\lambda$ is in $\mu$m.
Here, the three terms correspond to the far-UV extinction, the near-IR/visible component, and the 2175~\AA bump strength (controlled by $c_4$), respectively.

{The normalization is set by the formula itself: the numerical constants in Equation~(\ref{equ:1}) are the values taken by the three Drude denominators at the $V$ band, $\lambda_\mathrm{V}=0.55~\mu$m, so that the three terms sum identically to unity there for any $(c_1,c_2,c_3,c_4)$.}
{We therefore treat $A_\mathrm{V}$ as an explicit multiplicative free parameter, $A_\lambda^\mathrm{model}=A_\mathrm{V}\times\left[A_\lambda/A_\mathrm{V}\right](c_1,c_2,c_3,c_4)$, and fit it simultaneously with $c_1$, $c_2$, $c_3$, $c_4$ and the normalization $N_\mathrm{norm}$ of the intrinsic template introduced in Equation~(\ref{equ:2}) below.}
{The fitted $A_\mathrm{V}$ is thus by construction the extinction at 5500~\AA, and no external calibration or post-hoc rescaling is applied at any stage.}
{This fixes the definition of $A_\mathrm{V}$ but not its degeneracy with $N_\mathrm{norm}$; we quantify that degeneracy, and the resulting sensitivity of the inferred far-UV steepness, in Appendix~\ref{app:robust}.}

To determine the intrinsic flux, we utilize the QSO composite spectrum ($f_\mathrm{\lambda,template}$) of \citet{Selsing2016A&A...585A..87S}, displayed in Figure~\ref{fig:dustlaw}.
We model the intrinsic QSO spectrum ($f_\mathrm{\lambda,int}$) by applying a power-law modification to this template to account for slope variations:
\begin{equation}
f_\mathrm{\lambda,int} = N_\mathrm{norm}f_\mathrm{\lambda,template}(\frac{\lambda}{\lambda_0})^{\alpha_0-\alpha_\lambda},
\label{equ:2}
\end{equation}
where $N_\mathrm{norm}$ is the flux normalization factor, $\lambda_0=1~\mathrm{\mu m}$, and the template slope is $\alpha_0=1.7$.
{The template is initially scaled to the observed spectrum by the median flux ratio over the reddest 25 PRISM pixels, rest-frame 0.950--0.980~$\mu$m (observed 5.28--5.45~$\mu$m, median signal-to-noise ratio $\approx50$), where the extinction is smallest; $N_\mathrm{norm}$ is thereafter a free parameter of the fit, so this scaling only sets the starting point and does not constrain the result.}
We explore intrinsic slope values in the range $\alpha_\lambda=1.3-1.7$, which encompasses the intrinsic continuum slopes typically observed in QSOs (see their Table~2 in \citealt{Selsing2016A&A...585A..87S}).
We derive the extinction parameters by fitting the observed flux ratio defined as:
\begin{equation}
A_\lambda=-2.5\log(\frac{f_\lambda}{f_\mathrm{\lambda,int}}).
\label{equ:3}
\end{equation}

{Equation~(\ref{equ:3}) assumes that the dust acts as a uniform foreground screen.}
{The relevant scales differ by orders of magnitude.}
{For $M_\mathrm{BH}=2.5\times10^{8}\,M_\odot$ and an Eddington ratio of 0.5, the standard thin-disc relation places the UV--optical continuum-emitting region within $R\lesssim3\times10^{15}$~cm \citep{Shakura1973A&A....24..337S}, and microlensing measurements indicate that real discs are larger still by a factor of a few \citep{Morgan2010ApJ...712.1129M}.}
{Dust, by contrast, survives only outside the sublimation radius, for which near-infrared reverberation gives $R_\mathrm{sub}\approx0.5\,(L_\mathrm{bol}/10^{46}\,\mathrm{erg\,s^{-1}})^{1/2}\approx0.6$~pc \citep{Koshida2014ApJ...788..159K}, itself a factor of $\sim3$ inside the classical sublimation radius of \citet{Barvainis1987ApJ...320..537B}; we adopt the smaller, more conservative value.}
{The obscuring material therefore lies at least two orders of magnitude further from the black hole than the entire continuum-emitting region.}
{Whether a given structure at that distance covers the disc uniformly nevertheless depends on its transverse size, which is not directly constrained, so we test the screen assumption empirically rather than assume it.}
{First, the broad lines and the continuum are reddened by the same column: the observed rest-frame equivalent widths of C\,{\sc iv}, C\,{\sc iii}], Mg\,{\sc ii}, and H$\alpha$ agree with those of the intrinsic composite to within a factor of $\lesssim1.7$, even though the continuum transmission varies by a factor of 112 across this range, whereas a geometry in which the broad-line region escaped the obscuration would have boosted the C\,{\sc iv} equivalent width by a factor of 189 \citep[cf.][]{Gaskell2004ApJ...616..147G}.}
{Second, fitting an explicit covering fraction $f_\mathrm{c}$ drives it to unity, bounding any additive unobscured or scattered component at $1-f_\mathrm{c}<0.007$ ($3\sigma$), while a partially covering SMC or Milky Way law is strongly excluded.}
{Both a leaky and a clumpy screen \textit{flatten} rather than steepen the emergent curve, so our measured extinction curve is a conservative lower limit on the steepness of the underlying grain extinction law.}
{These tests are presented in full in Appendix~\ref{app:robust}.}

In our initial fit, we allow $c_1$, $c_2$, $c_3$, $c_4$, $A_\mathrm{V}$, and $N_\mathrm{norm}$ to vary freely, while treating $\alpha_\lambda$ as a grid parameter with a step size of 0.05.
Wavelengths bluer than Ly$\alpha$ are excluded, and those containing bright emission lines are masked during the fitting process.
{We adopt a uniform uncertainty of 0.1~mag on $A_\lambda$ rather than the formal propagated errors (median 0.013~mag), since the error budget is dominated by the intrinsic QSO-to-QSO dispersion about the composite rather than by photon noise.}
{The results yield a bump strength consistent with zero, $c_4=-0.007_{-0.009}^{+0.008}$.}
{We quantify the bump as the peak excess of the Drude term over the bump-free curve at 2175~\AA, which in this formalism is $A_\mathrm{bump}/A_\mathrm{V}=19.55\,c_4$, so that the free fit corresponds to $A_\mathrm{bump}/A_\mathrm{V}=-0.14_{-0.18}^{+0.15}$.}
{Adopting the physical prior $c_4\ge0$, we obtain upper limits of $c_4<0.0046$ ($A_\mathrm{bump}/A_\mathrm{V}<0.09$) at $1\sigma$ and $c_4<0.0172$ ($A_\mathrm{bump}/A_\mathrm{V}<0.34$) at $3\sigma$.}
{For reference, the Drude fits to the standard extinction laws tabulated by \citet{Li2008ApJ...685.1046L} correspond to $A_\mathrm{bump}/A_\mathrm{V}=1.01$ for the Milky Way ($c_4=0.0519$), 0.43 for the LMC ($c_4=0.0221$) and 0 for the SMC ($c_4=0$).}
{An injection--recovery experiment (Appendix~\ref{app:robust}) shows that the recovery is linear and unbiased, that the $3\sigma$ detection threshold for PRISM data of this quality is $A_\mathrm{bump}/A_\mathrm{V}\approx0.5$, and that a Milky-Way-strength bump would have been recovered at $5.1\sigma$; the sensitivity is set by the covariance between $c_4$ and the smooth far-UV terms rather than by spectral resolution, since the 486~\AA\ rest-frame width of the profile spans $\approx15$ PRISM pixels at the redshifted bump position.}
{The absence of the 2175~\AA\ feature is therefore a measurement rather than a non-detection driven by insufficient sensitivity, as is also apparent from the empirical extinction law (deep red line) in Figure~\ref{fig:dustlaw}.}
{Since $c_4$ is consistent with zero and unconstrained on the negative side, we set $c_4=0$ for the final analysis and reoptimize the model with the remaining five free parameters.}
The fitting parameters are $c_1=2.37_{-0.48}^{+0.68}\times10^3$, $c_2=-6.28\pm0.14$, $c_3=357_{-46}^{+53}$, and $A_\mathrm{V}=0.72\pm0.15$.

We apply the best-fit dust extinction model to de-redden the observed spectra (Fig.~\ref{fig:dustlaw}, left panel), revealing good agreement with the template continuum outside of wavelength ranges of major emission lines (illustrated for $\alpha_\lambda=1.5$).
The resulting extinction curve ($A_\lambda/A_\mathrm{V}$ \textit{v.s.} $1/\lambda$) is shown in the right panel of Fig.~\ref{fig:dustlaw}, including the $1\sigma$ uncertainty band derived from the MCMC posterior.
The extinction toward \target is distinctive when compared to standard benchmarks, including the Milky Way \citep{Cardelli1989ApJ...345..245C}, SMC \citep{Gordon2003ApJ...594..279G} extinction curves, the \citet{Calzetti2000ApJ...533..682C} starburst attenuation laws, and the attenuation curves inferred for $z>4$ galaxies observed by JWST/NIRSpec \citep{Markov2025NatAs...9..458M}.
We observe moderate optical {extinction} of $A_\mathrm{V}=0.72\pm0.15$ with a canonical total-to-selective extinction ratio $R_\mathrm{V}\equiv\frac{A_V}{A_B-A_V}=3.13_{-0.10}^{+0.07}$, where $A_B$ is the B-band (0.44~$\mu$m) dust extinction.
However, the extinction curve exhibits a significantly steeper rise into the UV, with a measured effective UV-optical slope (steepness) of $S\equiv A_{1500\AA}/A_\mathrm{V}=8.0_{-0.9}^{+1.4}$.
{At fixed intrinsic slope the statistical uncertainty on $A_\mathrm{V}$ is only 0.02~mag; the quoted error is dominated by the marginalization over $\alpha_\lambda=1.3$--$1.7$, which shifts $A_\mathrm{V}$ monotonically from 0.53 to 0.91 and $S$ from 10.1 to 6.9 (Appendix~\ref{app:robust}).}
{The steepness therefore exceeds $S>6.8$ for any intrinsic slope within the range spanned by observed QSOs.}
This steep slope, combined with moderate V-band {extinction}, positions the dust properties of \target\ at the extreme end of the currently known distribution (Fig.~\ref{fig:dustmodel}, left panel).

\begin{figure*}[t]
\centering
\includegraphics[width=\linewidth]{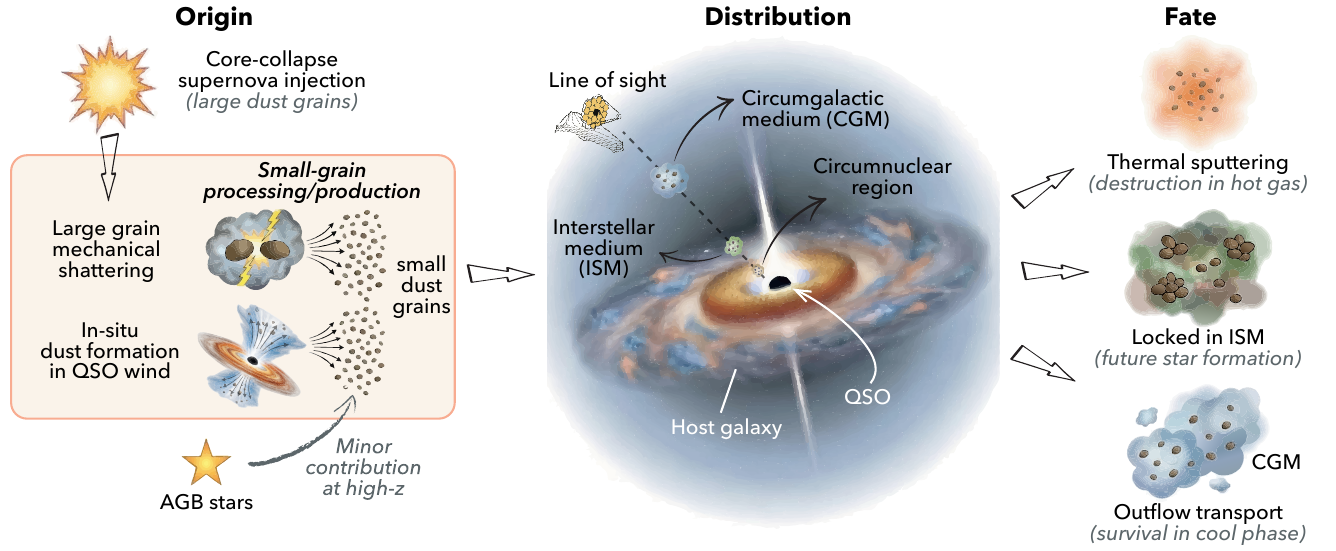}
\caption{
Schematic illustration of the origin, distribution, and fate of small dust grains in the steep-extinction QSO. \textit{Left (Origin):} In the early Universe, dust is predominantly injected as large grains by core-collapse supernovae (CCSNe), with minor contributions from AGB stars. Small grains may then be produced through two rapid channels in the QSO environment. In the first pathway, violent feedback drives grain--grain collisions, mechanically shattering large grains into small fragments. In the second pathway, silicate grains may condense in situ within dense QSO-driven winds. \textit{Middle (Distribution):} The JWST line of sight probes through the circumnuclear region, the host galaxy's interstellar medium (ISM), and the circumgalactic medium (CGM), directly detecting the steep extinction signature of an abundant small-grain population against the background QSO continuum. \textit{Right (Fate):} These small grains face three primary evolutionary pathways: they may be destroyed via thermal sputtering in hot, shock-heated gas; they may remain locked in the ISM to facilitate subsequent molecular-cloud and star formation; or they may be entrained in cool-phase outflows and transported to the CGM, populating the extended, small-grain-rich halos recently observed around massive galaxies at high redshift.}
\label{fig:cartoon}
\end{figure*}

\section{Discussion} \label{sec:discussion}

\subsection{Abundant small dust grains}
\label{subsec:smallgrains}

The extinction curve of \target is remarkable for its extreme steepness ($S=8.0_{-0.9}^{+1.4}$) and the absence of the 2175~\AA\ bump.
To interpret the physical properties of the dust responsible for this signature, we compare the observed extinction curve with theoretical models based on the MRN grain-size distribution \citep{Mathis1977ApJ...217..425M}, assuming a power-law distribution $\mathrm{d}n/\mathrm{d}a\propto a^{-q}$ with varying slope indices $q$ (Fig.~\ref{fig:dustmodel}, right panel).
Standard dust models for the Milky Way typically require $q=3.5$, $a_\text{min}=0.005\,\rm \mu m$, and $a_\text{max}=0.25\,\rm \mu m$ to reproduce the observed balance of optical and UV extinction.
As shown in the right panel of Fig.~\ref{fig:dustmodel}, a standard Milky Way mixture (solid grey curve) and a flatter extinction curve given by $q=2.5$ (blue dashed curve) are far too flat to reproduce the UV rise observed in \target.
Instead, the data are well qualitatively matched by a significantly steeper size distribution with $q=3.8$ (dashed black curve).
In the Rayleigh regime ($a\ll \lambda/2\pi$), absorption is determined by the total volume of dust, but in the UV, extinction becomes highly sensitive to the surface area provided by small grains.
The steep power-law index $q=3.8$ implies a grain mass distribution heavily skewed toward very small sizes ($a\sim0.01~\rm \mu m$), providing the necessary UV opacity without over-attenuating the optical continuum.

Furthermore, the absence of the 2175~\AA\ feature imposes a critical constraint on grain composition.
This feature is widely attributed to $sp^{2}$-bonded carbonaceous grains \citep{Draine1984ApJ...285...89D}, such as graphite or PAHs.
Its absence in \target is consistent with a grain population dominated by small-grain silicates, which exhibit a smooth, featureless rise into the far-UV.

This phenomenology stands in stark contrast to recent statistical studies of high-redshift galaxies.
For instance, \citet{Markov2025NatAs...9..458M} found that attenuation curves at $z>4$ are generally flatter than the SMC law, reflecting the dominance of large grains injected by core-collapse supernovae.
\target\ therefore could represent a distinct evolutionary phase or environment where the large-grain population has been fundamentally altered to abundant small dust grains.
{However, we caution against over-interpreting the comparison between attenuation and extinction curves, since a steep extinction curve integrated over a resolved, clumpy galaxy can emerge as a flat attenuation curve} \citep[e.g.,][]{Gallerani2010A&A...523A..85G, Salim2020ARA&A..58..529S}{.}

{It is therefore important to emphasize that what we measure toward \target is a genuine line-of-sight extinction curve rather than an attenuation curve.}
{Two mechanisms flatten attenuation curves: scattering of photons back into the beam, and the superposition of a distribution of dust columns across a spatially extended source \citep{Witt2000ApJ...528..799W, Seon2016ApJ...833..201S, Narayanan2018ApJ...869...70N}.}
{Both are strongly suppressed here.}
{The continuum arises from a single sub-parsec accretion disc that is unresolved in every NIRCam band (Fig.~\ref{fig:imgspec}), so there is no differential obscuration across an extended source, which removes the dominant mechanism that flattens galaxy attenuation curves.}
{In-beam scattering is not excluded by the aperture, however: the NIRSpec micro-shutter subtends $0\farcs20\times0\farcs46$, or $1.3\times3.0$~kpc at $z=4.556$, so a scattering halo on circumnuclear or host-ISM scales would fall entirely within it.}
{The relevant constraint is instead our leaky-screen fit, which bounds any additive unobscured or scattered path at $<0.7\%$ of the \textit{intrinsic} continuum at $3\sigma$ (Appendix~\ref{app:robust}).}
{We emphasise that such a component is nevertheless important where the transmission is smallest: a $0.3\%$ leak accounts for $\sim60$--$70\%$ of the \textit{observed} flux at 1500~\AA, which is precisely why including it drives the inferred intrinsic law steeper rather than flatter.}
{Crucially, all of the departures from a homogeneous screen that remain available (a partially resolved or clumpy obscurer, in-beam scattering, or residual unobscured light) act to \textit{reduce} the observed far-UV rise relative to the intrinsic grain extinction law, since each adds a weakly attenuated path that saturates the emergent curve at short wavelengths \citep{Natta1984ApJ...287..228N}.}
{A radiative-transfer model with a more realistic clumpy geometry therefore cannot reproduce the observed spectral energy distribution with a flatter underlying extinction law; it would require a steeper one, and hence an even more extreme small-grain dominance.}
{The value $S=8.0_{-0.9}^{+1.4}$ should accordingly be regarded as a conservative lower limit.}

{This argument, however, applies specifically to a foreground screen, in which any additional light path is weakly attenuated and therefore flattens the emergent curve.}
{If instead the dust is local to the source, multiple scattering can act in the opposite sense: photons are scattered repeatedly before escaping, with the effective path length increasing toward shorter wavelengths, so that the \textit{effective} extinction law becomes steeper than the underlying grain law even for standard grain sizes \citep{Goobar2008ApJ...686L.103G}.}
{\citet{Leighly2014ApJ...788..123L} showed that such a scattering-dominated law reproduces the anomalously steep reddening of Mrk~231, and \citet{Krawczyk2015AJ....149..203K} found that this ``L14-QSO'' curve describes the reddening of a subset of Type~1 quasars about as well as an SMC law.}
{This is a genuine alternative to a small-grain-dominated size distribution, and a single epoch of unpolarized spectroscopy cannot exclude it.}
{We note only that the mechanism characteristically produces a low effective $R_\mathrm{V}$, whereas the optical part of our curve is canonical ($R_\mathrm{V}=3.13_{-0.10}^{+0.07}$) and the extreme behaviour is confined to the far-UV; spectropolarimetry, which would detect the scattered component directly, offers the cleanest discriminant.}
{We note, however, that a single unresolved sightline cannot constrain the three-dimensional dust distribution or disentangle the circumnuclear, interstellar, and circumgalactic contributions to the column; spatially resolved integral-field spectroscopy and multi-epoch monitoring will be required for that.}

A complementary view emerges from low-ionization broad absorption line (LoBAL) and iron LoBAL (FeLoBAL) QSOs, which frequently exhibit steep reddening, with some of them accompanied by prominent 2175~\AA\ bumps \citep[e.g.,][]{Hall2002ApJS..141..267H, Jiang2013AJ....145..157J, Meusinger2016A&A...587A..83M, Krogager2016ApJ...832...49K, Zhang2022A&A...663A..63Z}.
A similar steep extinction without prominent bumps has been reported in larger samples of reddened QSOs \citep{Zafar2015A&A...584A.100Z} and in narrow-line Seyfert 1 galaxies \citep{Zhou2006ApJS..166..128Z}, which are usually reproduced by silicate-dominated small dust grains.
{Steep nuclear reddening is likewise seen in individual objects hosting powerful outflows, such as the narrow-line Seyfert~1 WPVS~007 \citep{Leighly2009ApJ...701..176L} and the BAL quasar SDSS~J1352$+$4239 \citep{Choi2020ApJ...891...53C}, reinforcing the association between outflow activity and anomalous reddening.}
{We stress that this is not in conflict with the flat, small-grain-deficient nuclear extinction curve derived by \citet{Gaskell2004ApJ...616..147G}: that curve is an ensemble average over many AGN and is dominated by moderately reddened sightlines, whereas \target, like the individual reddened QSOs and narrow-line Seyfert 1 galaxies cited above, represents a single extreme sightline.}
{The existence of both populations is precisely the point: nuclear dust spans a wide range of grain-size distributions, and SEQs occupy its small-grain extreme.}
The bump features are interpreted as small carbonaceous grains (graphite/PAHs) surviving in the nuclear region, protected by dense outflow clouds that shield them from the intense AGN UV/X-ray radiation {\citep{Zhang2022A&A...663A..63Z}}.

The coexistence of steep extinction curves and BAL troughs in these systems indicates that powerful AGN-driven outflows play a central dual role: they shatter large grains into small silicates (producing the featureless, ultra-steep UV extinction) while simultaneously shielding fragile carbonaceous grains in denser clumps (allowing 2175~\AA\ bumps to survive).
However, for \target, the relatively low spectral resolution of our NIRSpec PRISM data ({$R\approx35$--60 over the rest-frame UV, adopting two pixels per resolution element}) makes it challenging to securely identify or rule out the presence of BAL troughs.
Consequently, while \target appears as a classical Type~I QSO without evident BAL signatures, we cannot definitively classify it as a non-BAL steep-extinction QSO or as a LoBAL/FeLoBAL analog viewed through a shattering-active sightline.
Nevertheless, the available X-ray constraint disfavors the latter interpretation in its most classical, heavily obscured form.
The source has an X-ray-to-bolometric luminosity ratio consistent with the standard QSO relation, suggesting minimal X-ray absorption.
This contrasts with the strong X-ray weakness commonly observed in optically selected LoBAL/FeLoBAL QSOs, which is typically attributed to large absorbing columns reaching $  N_{\rm H}\sim10^{23}  $--$10^{24},{\rm cm^{-2}}$ in FeLoBALs \citep[e.g.,][]{Green2001ApJ...558..109G, Gallagher2006ApJ...644..709G, Rogerson2011NewA...16..128R, Morabito2011ApJ...737...46M}.
Thus, although an X-ray-bright BAL orientation cannot be ruled out, the X-ray detection makes a heavily X-ray-suppressed LoBAL/FeLoBAL analog less likely.
High-resolution follow-up spectroscopy (e.g., with the NIRSpec grating, ground-based optical/near-IR spectrographs, or future IFU observations), together with deeper X-ray data sufficient to constrain $\alpha_{\rm OX}$ and intrinsic absorption, will be essential for resolving this ambiguity.

A particularly relevant low-redshift analog is the UV-cutoff QSO SDSS J2317+0005 at $z=0.32$, reported by \citet{Guo2016ApJ...826..186G}.
In that object, the rest-frame UV continuum at $\sim3000\,{\rm \AA}$ suddenly dimmed by a factor of $\sim3.5$ on a timescale of only 23 days, while the broad emission-line fluxes remained nearly unchanged, and the source appeared to recover within the following several tens of days.
The extinction curve inferred from the dim state is substantially steeper than the SMC law, and was interpreted as a possible eclipse by a dusty cloud, potentially associated with a rapid inflow or outflow crossing the line of sight to the central engine.
Although \target is currently observed only in a single spectroscopic epoch, the close phenomenological similarity suggests that SEQs may not necessarily represent a long-lived quasar subtype.
Instead, they may mark a very short-lived transiting phase, during which a compact, small-grain-rich dusty structure temporarily intercepts the line of sight to the accretion disk.
If the duty cycle of this phase is indeed as short as days to months in the rest frame, such objects would be intrinsically rare in single-epoch spectroscopic surveys.
{We note that the timescales set a consistency requirement: at the Keplerian speed of $\sim1300$~km~s$^{-1}$ at $R_\mathrm{sub}$, a structure large enough to cover the continuum source uniformly would take $\sim10$--$100$~yr to transit, whereas a days-to-months eclipse of the kind seen in SDSS~J2317+0005 requires an occulter comparable to or smaller than the disc itself.}
{The latter is also the one geometry that could in principle steepen the emergent curve without invoking small grains, since the disc emitting radius scales as $R\propto\lambda^{4/3}$, with $R(5500~\mathrm{\AA})/R(1500~\mathrm{\AA})=5.6$, so that an occulter of intermediate size would cover the far-UV emitting region more completely than the optical one.}
{Such a configuration would, however, decouple the broad-line region from the continuum reddening and produce a saturating blue end, both of which are excluded by the tests of Appendix~\ref{app:robust}; repeat spectroscopy searching for the variability that this scenario predicts would settle the question directly.}
Systematic JWST/NIRSpec searches for SEQs, together with repeat NIRSpec spectroscopy and multi-band NIRCam monitoring, will therefore be essential for measuring their occurrence rate, lifetime, redshift evolution, and connection to QSO-driven dust processing.

While the current sample of steep-extinction QSOs (SEQs) at high redshift ($z>3$) remains limited, \target is not an isolated case.
This object was identified through visual inspection of a small sample of luminous QSOs observed in JWST/NIRSpec programs.
In addition to \target, at least one other SEQ has been identified at $z=3.71$ (ID 209777 in JWST-GTO-1180), though a detailed analysis is beyond the scope of this work.
These discoveries suggest that a population of such SEQs exists at high redshifts, potentially representing a common yet previously under-recognized evolutionary phase in the dust processing of early AGN hosts.

\subsection{Origin, distribution, and fate of small dust grains} 

{Before turning to the lifecycle of these grains, it is instructive to place our measurement in the context of current models of dust formation and processing in the high-redshift Universe \citep[for a review, see][]{Schneider2024A&ARv..32....2S}, and to compare it in detail with the SMC, the closest local analogue of a bumpless, steep extinction curve.}
{The contrast with the 2175~\AA\ bump detected at $z=6.7$ by \citet{Witstok2023Natur.621..267W}, which established that bump carriers can already be present within the first billion years \citep[see also the discussion of][]{Yang2023Natur.621..260Y}, is particularly instructive: the two measurements together indicate that the carbonaceous carrier is present in some early environments and absent in others, rather than simply unformed at early times.}
{The SMC Bar law is the closest local match in \textit{character}: it lacks the 2175~\AA\ bump and rises steeply into the UV along most sightlines, behaviour usually attributed to low metallicity, a low C/O ratio, and efficient destruction of carbonaceous carriers by the ambient radiation field \citep{Gordon1998ApJ...500..816G, Gordon2003ApJ...594..279G}.}
{It is, however, far from a match in \textit{degree}.}
{The Drude fit to the SMC law tabulated by \citet{Li2008ApJ...685.1046L} gives $S=4.46$ with no bump, to be compared with $S=8.0_{-0.9}^{+1.4}$ and $A_\mathrm{bump}/A_\mathrm{V}<0.09$ ($1\sigma$) for \target.}
{The extinction toward \target is thus 1.8 times steeper than the SMC and 3.1 times steeper than the Milky Way ($S=2.60$ in the same formalism), corresponding to $q\approx3.8$ rather than the MRN value $q=3.5$ that approximately reproduces both local laws.}
{Low metallicity and a carbon-poor composition alone, the standard SMC recipe, are therefore not sufficient to account for what we observe; an actively operating small-grain production or processing mechanism is required.}

Building upon this diversity of small-grain production and processing channels in luminous AGN, Fig.~\ref{fig:cartoon} summarizes the possible lifecycle of the small silicate grains that dominate the line-of-sight extinction toward UDS-27023.
One natural pathway begins with large grains ($a \gtrsim 0.1\,\mu{\rm m}$) injected by core-collapse supernovae on short timescales ($\lesssim10$~Myr; \citealt{Todini2001MNRAS.325..726T, Maiolino2004Natur.431..533M, Gall2011A&A...528A..13G}).
Within the dense, turbulent ISM and circumnuclear environment of this luminous QSO host, powerful shocks and outflows driven by the central engine can induce grain--grain collisions.
These collisions efficiently shatter large grains into small fragments \citep{Hirashita2009MNRAS.394.1061H, Hirashita2010MNRAS.407L..49H, Hirashita2013EP&S...65.1083H}, generating a silicate-dominated population of very small grains ($a\sim0.01\,\mu{\rm m}$) that reproduces the exceptionally steep, featureless far-UV extinction observed along the JWST sightline.
This mechanical pathway dramatically increases the total grain surface area available for accretion, enabling rapid grain growth in the ISM and helping to alleviate the long-standing dust-budget crisis at high redshift \citep{Kuo2012MNRAS.424L..34K, Asano2013MNRAS.432..637A}.

A further possibility, not mutually exclusive with mechanical shattering, is that part of the small-grain population forms \textit{in situ} within the QSO-driven wind itself.
This ``smoking quasar'' scenario was originally proposed by \citet{Elvis2002ApJ...567L.107E}, in which dense broad-line clouds embedded in an outflow expand and cool until they enter the dust-formation window.
More detailed calculations by \citet{Sarangi2019ApJ...885..126S} show that AGN accretion-disk winds can provide favorable conditions for silicate dust formation over parsec scales, especially in luminous systems accreting at a substantial fraction of the Eddington limit.
Importantly, these models predict distinct regions within the flow: in some zones, seed nuclei can form but do not efficiently accrete or coagulate, while in others they grow into silicate grains.
The freshly formed dust is therefore expected to be dominated by small silicate grains, with some grains remaining at sizes on the order of a few hundred Angstroms or below before further growth or processing.
This provides an intriguing additional pathway for producing the very small, silicate-dominated grains inferred in \target.
Given its large black-hole mass and high Eddington ratio, \target may therefore be probing a phase in which QSO feedback both processes pre-existing grains through shocks and produces new dust directly within the outflow.
Current data constrain the grain population but do not uniquely distinguish between these two channels; high-resolution spectroscopy capable of identifying outflow/BAL signatures, together with future constraints on the infrared-to-submillimeter dust SED, will be required to determine their relative importance.

{These two channels also account for the composition we infer, and in particular for the question of why, under the physical conditions traced here, small silicate grains should form and persist while carbonaceous grains do not.}
{We identify four mutually reinforcing reasons.}
{First, the supply is O-rich.}
{In CCSN ejecta, carbonaceous grains condense only in the thin, C-rich outer zones, whereas silicates and oxides condense throughout the far more massive O/Mg/Si zones: in equilibrium condensation calculations this is because CO sequesters carbon in the oxygen-rich material \citep{Todini2001MNRAS.325..726T, Nozawa2003ApJ...598..785N, Bianchi2007MNRAS.378..973B, Sarangi2015A&A...575A..95S}, while in non-equilibrium chemical-kinetic treatments carbon dust is instead restricted by the destruction of carbon-chain precursors by He$^+$ \citep{Cherchneff2010ApJ...713....1C}.}
{Either way, the silicate-forming zones dominate by mass, so a supernova-dominated inventory is silicate-weighted from the outset.}
{Second, the in-situ channel is likewise known to produce silicates: the chemical-kinetic calculations of \citet{Sarangi2019ApJ...885..126S} show that AGN accretion-disc winds in luminous, near-Eddington systems provide favourable conditions for \textit{silicate} nucleation and growth over parsec scales.}
{We note that an equivalent carbon-condensation channel in such winds has not, to our knowledge, been computed, so this argument bears on what is known to form rather than on what is excluded.}
{Both of the production channels described above therefore deliver silicates preferentially.}
{Third, the bump carrier is selectively destroyed.}
{The 2175~\AA\ feature is attributed to small $sp^2$-bonded carbonaceous material, that is, nano-graphite and/or PAHs, although the precise carrier remains debated \citep{Draine1984ApJ...285...89D, Lin2023MNRAS.525.2380L}.}
{This is the most fragile dust component in an AGN environment, being photo-dissociated and Coulomb-exploded by the intense EUV/X-ray radiation field \citep{Voit1992MNRAS.258..841V}.}
{Consistently, the short-wavelength PAH bands carried by the smallest molecules are suppressed in AGN-dominated nuclei \citep{Smith2007ApJ...656..770S}, and the PAH population that survives within the inner kiloparsec of Seyferts is shifted toward large, neutral species \citep{Garcia-Bernete2022MNRAS.509.4256G}; that is, precisely the small carriers responsible for the bump are the ones preferentially removed.}
{Silicates have no comparable photo-destruction channel and are removed mainly by thermal sputtering, which acts on both species and on longer timescales.}
{Importantly, the destroying radiation field is directly measured in this source, $\log(L_\mathrm{2-10\,keV}/\mathrm{erg\,s^{-1}})=44.42$ with a normal bolometric correction, indicating a sightline that is not heavily shielded.}
{Fourth, shattering conserves composition: grain--grain collisions fragment large silicates into small silicates \citep{Hirashita2009MNRAS.394.1061H, Hirashita2010MNRAS.407L..49H}, but cannot build the ordered $sp^2$ nano-structure responsible for the 2175~\AA\ feature, which shocks and ion irradiation instead amorphize.}
{A population dominated by shattered supernova silicates and irradiated by a luminous, poorly shielded AGN is therefore expected to be both bumpless and strongly weighted toward small grains, exactly as observed.}
{We note that this argument concerns the relative yields and survival of the two species rather than the absolute carbon budget: AGB stars are already active dust producers by $z\simeq4.6$ \citep{Valiante2009MNRAS.397.1661V, Schneider2024A&ARv..32....2S}, but any carbonaceous grains they contribute are subject to the same nuclear radiation field.}

{This picture makes a testable prediction that also connects to the LoBAL/FeLoBAL comparison discussed in Section~\ref{subsec:smallgrains}.}
{The reddened QSOs in which 2175~\AA\ bumps have been reported are predominantly LoBAL and FeLoBAL systems, where the carriers are thought to survive because dense outflow clouds shield them from the nuclear radiation field \citep{Zhang2022A&A...663A..63Z}; these are also the systems in which large absorbing columns and X-ray weakness are common \citep[e.g.,][]{Green2001ApJ...558..109G, Gallagher2006ApJ...644..709G, Morabito2011ApJ...737...46M}.}
{To our knowledge these two properties have not been tested against each other, and we therefore offer the connection as a prediction rather than an established result: if shielding is indeed what preserves the bump carriers, bump strength in reddened QSOs should anti-correlate with the transparency of the sightline to the ionizing continuum, with \target, which is X-ray normal and bumpless, lying at one end of that sequence.}
{This is directly testable with existing samples of reddened and BAL QSOs having both UV extinction curves and X-ray constraints.}

Once produced, either through mechanical shattering or in-situ condensation within the QSO wind, these small grains follow three primary evolutionary pathways.
(1) In hot, post-shock gas ($T\gtrsim10^6$~K), they are rapidly destroyed by thermal sputtering on timescales of $\sim10^4$--$10^5$~yr \citep{Draine1979ApJ...231...77D, Jones1996ApJ...469..740J}.
(2) In shielded, dense molecular clouds, they survive and grow by accretion, fueling further dust-mass assembly and contributing to the substantial cold-dust reservoir detected at submillimeter wavelength \citep{Hirashita2012MNRAS.422.1263H}.
(3) Some fraction may be entrained in cool-phase outflows, ubiquitous around high-$z$ QSOs, and expelled into the CGM \citep{Cicone2018NatAs...2..176C, Laha2021NatAs...5...13L}.
Radiation pressure and galaxy interactions/mergers are therefore more naturally interpreted as redistribution mechanisms acting after small grains are produced, selectively exposing unshielded grains to destruction while allowing grains embedded in dense clumps to survive.
This channel could directly populate the extended reservoirs of small silicate grains recently revealed by JWST/NIRCam grism spectroscopy around $z\sim3.5$ massive galaxies, where steep extinction of background sources at 7--30~kpc indicates high dust surface densities ($\gtrsim10^{-1}\,M_\odot$~pc$^{-2}$, $\sim10\times$ local CGM values) and solar-like enrichment \citep{Sun2026arXiv260115961S}.
Shattering within turbulent cool CGM clumps can further amplify the small-grain population on timescales of a few $10^8$~yr \citep{Hirashita2021MNRAS.505.1794H}.

In this way, \target\ exemplifies how AGN at high redshift can play a dual role in the dust lifecycle: they may both process pre-existing grains through shocks and turbulence, and produce new dust directly within dense QSO-driven winds.
The resulting small grains can subsequently grow in the ISM or be redistributed into the circumgalactic medium.
Steep-extinction QSOs therefore bridge supernova dust injection, interstellar dust processing, QSO-wind dust formation, and halo enrichment, thereby closing the observational loop between the massive dust reservoirs seen in early galaxies and the small-grain-rich halos that regulate future star formation.

\section{Summary}\label{sec:summary}

In this work, we report the discovery of a steep-extinction QSO (SEQ) \target at $z=4.556$ using JWST/NIRSpec spectroscopy, revealing an exceptionally steep far-UV extinction curve ($S\equiv A_{1500}/A_V=8.0_{-0.9}^{+1.4}$) without a detectable 2175~\AA\ bump {($A_\mathrm{bump}/A_V=-0.14_{-0.18}^{+0.15}$, hence $<0.09$ and $<0.34$ at $1\sigma$ and $3\sigma$, against 1.01 for the Milky Way in the same formalism)}.
{Because any clumpy or partially covering geometry flattens rather than steepens the observed curve, this steepness is a conservative lower limit on that of the underlying grain extinction law.}
This signature, combined with a moderate visual extinction $A_V=0.72\pm0.15$ and a silicate-dominated small-grain model ($q=3.8$ in the MRN-like distribution), indicates an abundance of very small dust grains ($a\sim0.01~\mu{\rm m}$) along the line of sight.
ALMA observations further confirm a substantial cold-dust reservoir ($M_d\approx4\times10^8\,M_\odot$), consistent with a host galaxy undergoing rapid dust growth.
We interpret this as evidence for an active small-grain production/processing phase driven by the QSO.
One natural mechanism is mechanical shattering, in which powerful shocks and outflows pulverize supernova-injected large grains into small silicates; an additional possibility is direct condensation of small silicate grains within the QSO wind itself.
These channels could alleviate tensions in the high-redshift dust-budget problem by enabling efficient ISM accretion and growth, while also enriching the CGM with small-grain reservoirs.
In addition to UDS-27023, at least one other SEQ has been identified at $z=3.71$ (ID 209777 in JWST-GTO-1180).
SEQs like UDS-27023 could represent a transitional evolutionary stage in luminous AGN, bridging supernova dust injection, interstellar dust processing, QSO-wind dust formation, and circumgalactic enrichment.
Future high-resolution spectroscopy will refine the BAL/outflow classification, while deeper infrared and submillimeter observations will help constrain the dust emission, grain composition, and duty cycle of this small-grain production phase.

\begin{acknowledgments}
{We thank Patrick B. Hall, Karen M. Leighly, and Struan D. Stevenson for helpful correspondence following the submission of this work.}
This work is supported by National Key R\&D Program of China (grant no. 2023YFA1605600).
This research is supported by National Natural Science Foundation of China (\#12525303) and Tsinghua University Initiative Scientific Research Program. This work is funded by New Cornerstone Science Foundation through the XPLORER PRIZE.
RM acknowledges support from the Science and Technology Facilities Council (STFC), by the European Research Council (ERC) through Advanced Grant 695671 ``QUENCH'', by the UK Research and Innovation (UKRI) Frontier Research grant RISEandFALL. RM also acknowledges support from a Royal Society Research Professorship grant.
This work is based in part on observations made with the NASA/ESA/CSA James Webb Space Telescope. The data were obtained from the Mikulski Archive for Space Telescopes at the Space Telescope Science Institute, which is operated by the Association of Universities for Research in Astronomy, Inc., under NASA contract NAS 5-03127 for JWST. These observations are associated with program \#1215, \#1837, and \#7814.
The authors acknowledge the PRIMER team led by PI (J. Dunlop) and the MINERVA team led by CoPIs (A. Muzzin, D. Marchesini, and K. Suess) for developing their observing program with a zero-exclusive-access period.
(Some of) The data products presented herein were retrieved from the Dawn JWST Archive (DJA). DJA is an initiative of the Cosmic Dawn Center (DAWN), which is funded by the Danish National Research Foundation under grant DNRF140.
All the JWST raw data used in this paper can be found in MAST: \dataset[10.17909/yscx-cd19]{http://dx.doi.org/10.17909/yscx-cd19}.
This paper makes use of the following ALMA data: ADS/JAO.ALMA\#2012.1.00326.S and ADS/JAO.ALMA\#2015.1.01074.S. ALMA is a partnership of ESO (representing its member states), NSF (USA) and NINS (Japan), together with NRC (Canada), MOST and ASIAA (Taiwan), and KASI (Republic of Korea), in cooperation with the Republic of Chile. The Joint ALMA Observatory is operated by ESO, AUI/NRAO and NAOJ.
The National Radio Astronomy Observatory is a facility of the National Science Foundation operated under cooperative agreement by Associated Universities, Inc.
\end{acknowledgments}





%
\facilities{JWST (NIRCam, NIRSpec), ALMA, CXO}



\appendix

\section{Photometric Misclassification before and after JWST} \label{app:photz}

Prior to the JWST spectroscopic confirmation, \target was consistently classified as a massive quiescent galaxy at photometric redshift $z_\mathrm{phot}\approx2$--3, often with evidence for an embedded infrared-active AGN \citep{Bruce2012MNRAS.427.1666B, Cowley2016MNRAS.457..629C, Patil2019ApJ...871..109P, Carnall2020MNRAS.496..695C, Kodra2023ApJ...942...36K}.
This mis-identification stemmed from the object's unusual photometric colors, which were interpreted under the assumption of standard galaxy templates without accounting for the extremely steep-extinction QSO (SEQ) nature revealed by NIRSpec.
{It persists in the most recent photometric selections: \citet{Stevenson2026MNRAS.545f2087S}, an updated analysis of the \citet{Carnall2020MNRAS.496..695C} sample using three HST and eight JWST/NIRCam bands, also select \target as a massive quiescent galaxy (ID PRIMER-UDS-116064; their Fig.~4).}
{Even a dozen photometric bands spanning the optical to the near-infrared therefore do not break this degeneracy without spectroscopy.}

\begin{figure*}[h]
\centering
\includegraphics[width=\linewidth]{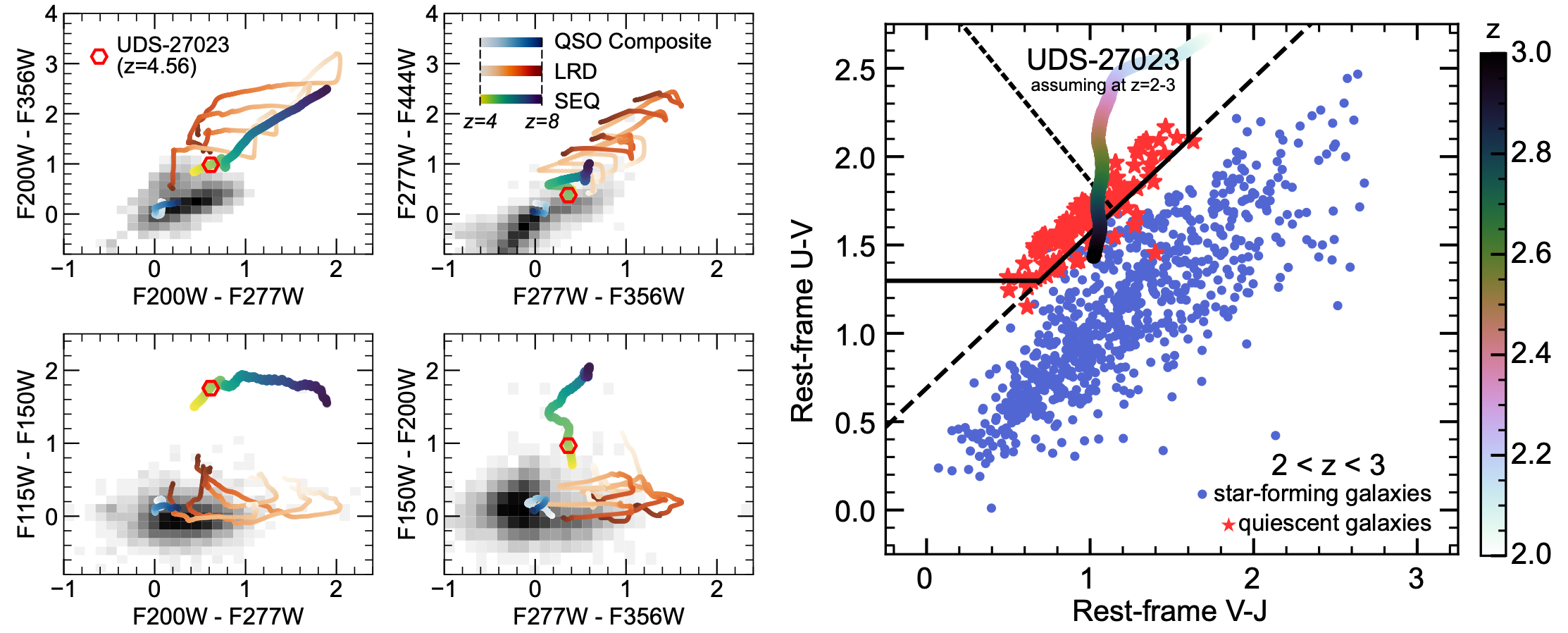}
\caption{
Photometric selection biases induced by SEQ-like extinction.
\textit{Left:} A set of illustrative JWST/NIRCam color--color diagrams demonstrating that \target\ (red symbol) lies outside common loci for AGN and QSOs in multiple color spaces. Colored tracks show synthetic colors for an intrinsic QSO composite \citep{Selsing2016A&A...585A..87S}, representative LRD-like colors \citep{Zhang2025arXiv251205180Z}, and the SEQ extinction model, evaluated as a function of redshift (color-coded). The grey background indicates the distribution of field sources from the JADES survey \citep{Robertson2026arXiv260115956R} in the same diagrams.
\textit{Right:} UVJ diagram adapted from Fig.~2 of \citet{Carnall2020MNRAS.496..695C}, illustrating possible misclassification of SEQ under an incorrect low-redshift interpretation.
We overlay the inferred UVJ track of \target when forced to $z=2$--3 (color-coded by assumed redshift) in galaxy-only SED fitting.
It falls within the standard UVJ-quiescent selection region (within black solid lines), consistent with its previous identification as a candidate of $z\approx2.5$ quiescent galaxy.
}
\label{fig:color}
\end{figure*}

\begin{figure*}[h]
\centering
\includegraphics[width=\linewidth]{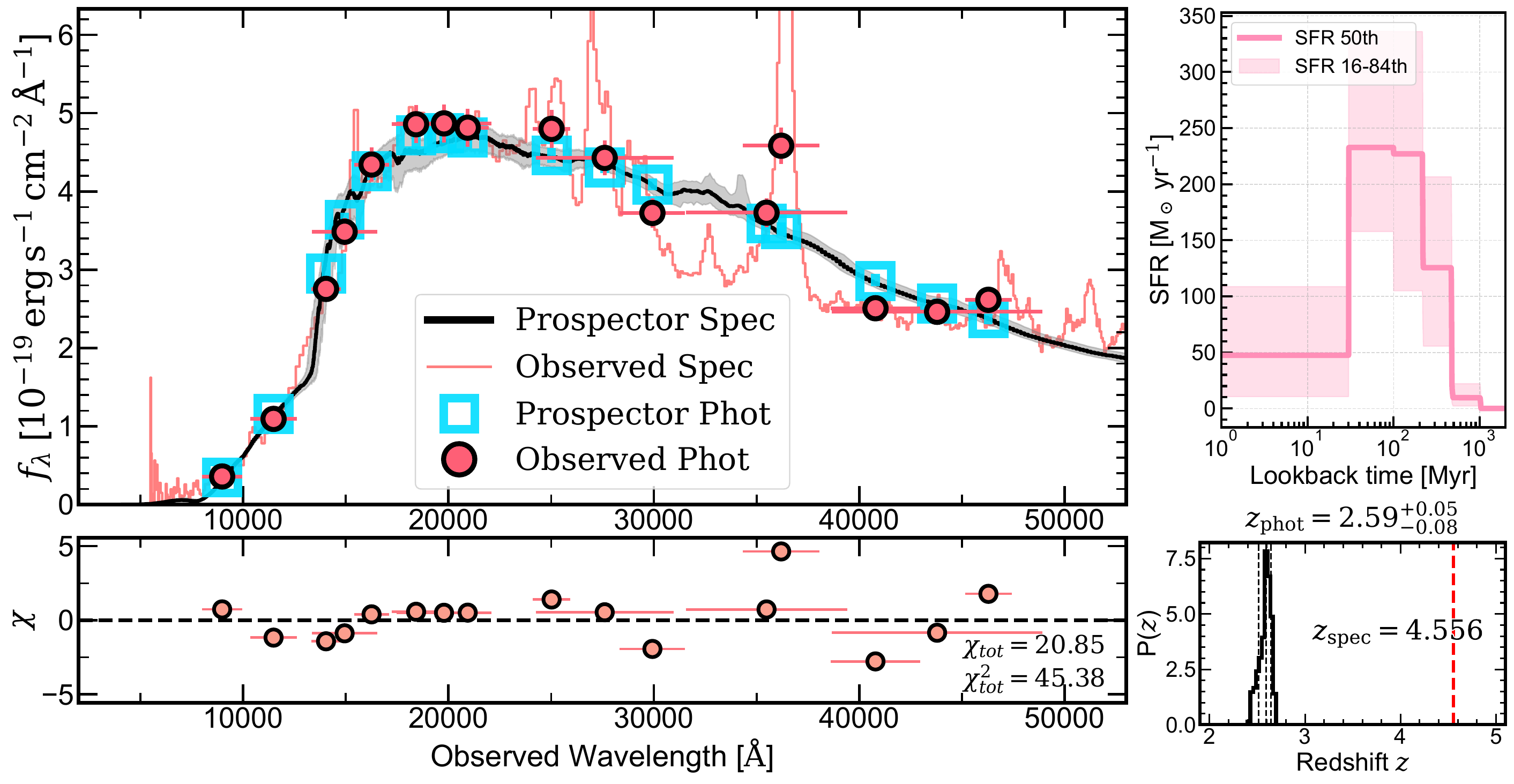}
\caption{
Photometric redshift and SED modeling on the JWST photometry for \target using the \texttt{Prospector} code, still revealing a quiescent/post-starburst galaxy at $z_\mathrm{phot}\approx2.6$.
\textit{Left:} Observed spectrum (thin red line) and photometry (filled red circles with error bars) compared with the best-fit Prospector model spectrum (thick black line) and synthetic photometry (open cyan squares).
Fit residuals $\chi$ are shown in the lower panel, with total $\chi^2$ values for the fit indicated.
\textit{Top right:} Posterior star-formation rate (SFR) as a function of lookback time.
\textit{Bottom right:} Photometric redshift probability distribution $P(z)$. The spectroscopic redshift $z_\mathrm{spec} = 4.556$ is marked by the vertical red dashed line.
}
\label{fig:SED}
\end{figure*}

Figure~\ref{fig:color} illustrates the photometric selection biases induced by SEQ-like extinction.
In the left panels, JWST/NIRCam color--color diagrams show that \target (red symbol) falls outside typical loci for AGN and QSOs.
Synthetic tracks for an intrinsic QSO composite \citep{Selsing2016A&A...585A..87S}, representative little red dot (LRD) colors \citep{Zhang2025arXiv251205180Z}, and our SEQ extinction model (evaluated as a function of redshift) demonstrate how the steep UV suppression shifts the object into regions mimicking lower-redshift dusty galaxies.
The grey background, drawn from sources in the JADES survey field, highlights this displacement.
The right panel adapts the UVJ diagram from \citet{Carnall2020MNRAS.496..695C}, showing the inferred UVJ track for \target when forced to $z=2$--3 in galaxy-only SED fitting (color-coded by assumed redshift).
This track lands squarely within the standard UVJ-quiescent selection region (black solid lines), aligning with prior classifications as a $z\approx2.5$ quiescent galaxy.

We further demonstrate that even photometric redshift and SED fitting of the new JWST photometry fails to yield a reliable (high-redshift) photometric redshift.
Following the approach of \citet{Duan2026arXiv260521599D}, we perform the fitting using the \texttt{Prospector} code \citep{Johnson2021ApJS..254...22J}, adopting a non-parametric star formation history \citep[SFH][]{Leja2019ApJ...876....3L, Tacchella2022ApJ...927..170T} with the star-forming main sequence (SFMS) prior, and allowing the redshift to vary freely.
We refer the reader to \citet{Duan2026arXiv260521599D} for the technical details of the parameters and priors used in the SED fitting.
Brief information is provided here.
The SFMS SFH prior is constructed by assuming that galaxies form and evolve along the SFMS with substantial scatter, and is applied as a prior on the non-parametric SFH.
In the non-parametric SFH framework, the SFH is discretized into different age bins, within which the SFR is allowed to vary.
The ratios between the SFRs in adjacent age bins are fitted as free parameters in the SED modeling, and the prior on these ratios follows a Student's $t$ distribution.
The SFMS prior modifies the expected value, $\mu$, of this Student's $t$ distribution to the value expected from the SFMS, while using $\nu = 2.0$ and $\sigma = 0.5$.
This means that the expected SFH is centered on the SFMS, while still allowing substantial scatter and deviations when supported by the data.
Therefore, if a galaxy is quiescent, the prior assigns a lower probability to this solution relative to an SFMS-like history.
However, if the galaxy is genuinely quiescent, as strongly supported by the data, this penalty term will not affect the posterior inference to any significant extent, and the posterior can still favor a data-driven quiescent solution.
Even with this state-of-the-art framework, the best-fit model (Fig.~\ref{fig:SED}) still classifies \target as a quiescent or post-starburst galaxy at $z_\mathrm{phot}\approx2.6$ ($z_\mathrm{phot}=2.59^{+0.05}_{-0.08}$).
We note that the derived star-formation history is unreliable due to an incorrect redshift.

The steep far-UV extinction effectively mimics the red colors of an old stellar population with moderate dust, while the QSO continuum boosts the rest-frame near-IR, further supporting an AGN-hosting quiescent interpretation.
This case underscores the challenges of photometric redshifts for reddened AGN without spectroscopic observations.
The JWST/NIRSpec spectrum resolves this ambiguity, confirming the true redshift and revealing the object’s genuine QSO nature.
Future surveys should incorporate flexible extinction models to mitigate such biases and assemble a more complete census of high-$z$ sources.

\section{{Robustness of the Extinction-Curve Fit}}
\label{app:robust}

{This appendix sets out the analysis underlying the results quoted in Section~\ref{sec:result} and Section~\ref{sec:discussion}, and is not intended to restate them: the decomposition of the $A_\mathrm{V}$ error budget, the derivation of the limits on the 2175~\AA\ bump together with an injection--recovery assessment of the detection threshold, and the tests of the foreground-screen assumption.}

\subsection{{Normalization and the degeneracies of $A_\mathrm{V}$}}

{The \citet{Li2008ApJ...685.1046L} parameterization is self-normalizing at the $V$ band, as stated in Section~\ref{sec:result}.}
{Explicitly, at $\lambda_\mathrm{V}=0.55~\mu$m one has $0.55/0.08=6.88$, $0.08/0.55=0.145$, $(0.55/0.046)^2+(0.046/0.55)^2+90=233$ and $(0.55/0.2175)^2+(0.2175/0.55)^2-1.95=4.60$, which are precisely the constants appearing in Equation~(\ref{equ:1}), so that the far-UV, visible and bump terms sum to unity there for any $(c_1,c_2,c_3,c_4)$.}
{We verify numerically that $A_{0.55\,\mu\mathrm{m}}/A_\mathrm{V}=1.0004$ across the full posterior, with a spread of only $7\times10^{-5}$; the residual $4\times10^{-4}$ offset arises solely from the rounding of those constants as published.}
{The $V$ band itself is directly observed, rest-frame 5500~\AA\ falling at $3.06~\mu$m where the PRISM spectrum has a signal-to-noise ratio of $\approx100$ per pixel, so $A_\mathrm{V}$ is not obtained by extrapolating the parametric form.}

{This fixes the definition of $A_\mathrm{V}$ but does not remove its degeneracies.}
{The reddest rest-frame wavelength sampled is 0.98~$\mu$m, where the best-fit law still gives $A_\lambda/A_\mathrm{V}=0.42$, so there is no wavelength in the data at which the extinction is negligible; $A_\mathrm{V}$ is correspondingly correlated with the template normalization at $r=0.97$, and the fit returns $N_\mathrm{norm}=1.29_{-0.06}^{+0.07}$.}
{An additional grey component would therefore be absorbed into $N_\mathrm{norm}$, raising $A_\mathrm{V}$ and lowering $S$: a grey $\Delta A=0.3$~mag would give $S=6.0$.}
{Our $S$ should accordingly be read as conservative with respect to clumping and partial covering, as established by the tests of the screen assumption below, but as an upper limit with respect to any grey extinction component.}

{We note also that $c_1$ and $c_3$ are strongly correlated in the posterior ($r=0.62$, with $c_1/c_3=6.6_{-1.1}^{+1.6}$), as are $c_2$ and $c_3$ ($r=-0.96$), so that it is the resulting curve rather than the individual coefficients that is constrained; the values of $c_1$--$c_3$ should not be compared directly with those tabulated for other extinction laws \citep[e.g.,][]{Li2008ApJ...685.1046L}.}

{The second systematic, which dominates the quoted uncertainty on $A_\mathrm{V}$, is the unknown intrinsic continuum slope $\alpha_\lambda$ of the QSO.}
{Table~\ref{tab:alphagrid} tabulates the posterior median of $A_\mathrm{V}$ and of the steepness $S$ at each point of the grid, from which the error budget quoted in Section~\ref{sec:result} follows.}

\begin{deluxetable}{lccccccccc}[t]
\tablecaption{{Dependence of the inferred extinction parameters on the assumed intrinsic QSO slope.}\label{tab:alphagrid}}
\tablehead{\colhead{{$\alpha_\lambda$}} & \colhead{{1.30}} & \colhead{{1.35}} & \colhead{{1.40}} & \colhead{{1.45}} & \colhead{{1.50}} & \colhead{{1.55}} & \colhead{{1.60}} & \colhead{{1.65}} & \colhead{{1.70}}}
\startdata
{$A_\mathrm{V}$ [mag]} & {0.53} & {0.58} & {0.62} & {0.67} & {0.72} & {0.77} & {0.82} & {0.87} & {0.91} \\
{$S\equiv A_{1500}/A_\mathrm{V}$} & {10.05} & {9.42} & {8.89} & {8.41} & {8.02} & {7.68} & {7.37} & {7.10} & {6.86} \\
\enddata
\tablecomments{{The statistical uncertainty at fixed $\alpha_\lambda$ is $\sigma(A_\mathrm{V})=0.023$~mag.
Marginalizing over the grid gives $A_\mathrm{V}=0.72\pm0.15$ and $S=8.0_{-0.9}^{+1.4}$.}}
\end{deluxetable}

\subsection{{Limits on the 2175~\AA\ bump}}

\begin{figure*}[t]
\centering
\includegraphics[width=\linewidth]{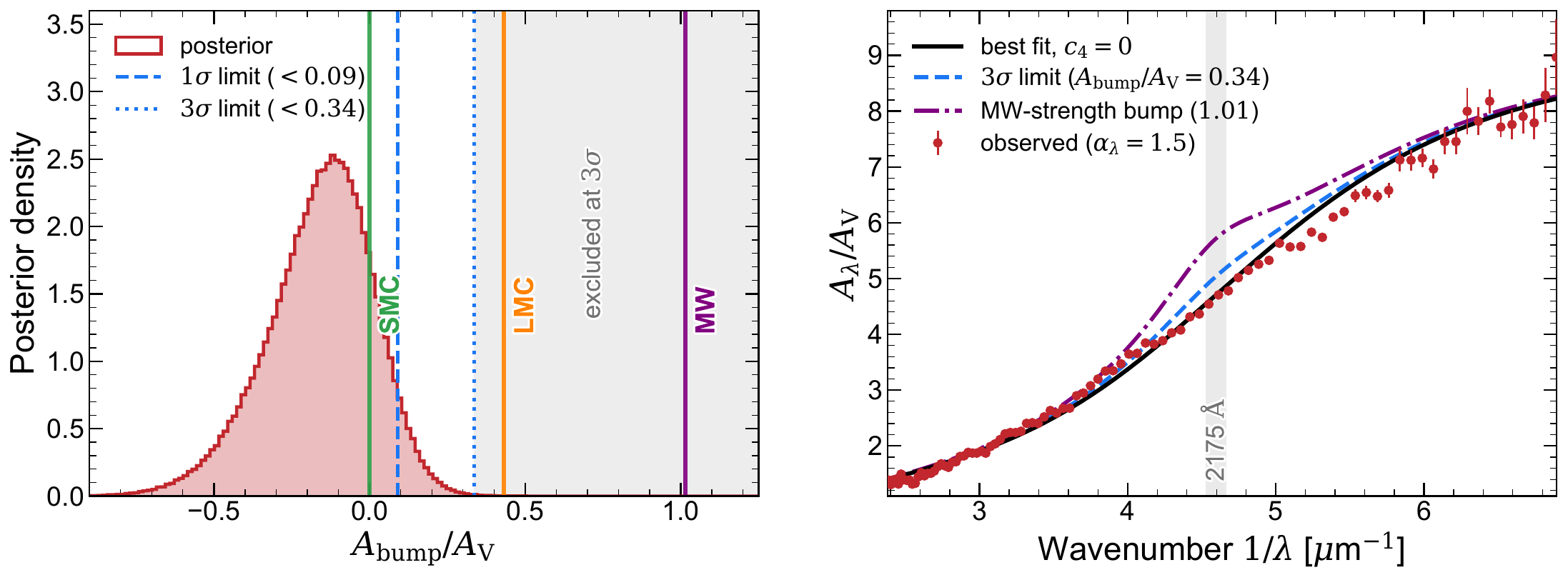}
\caption{{Constraints on the 2175~\AA\ bump toward \target.
\textit{Left:} posterior distribution of the bump strength $A_\mathrm{bump}/A_\mathrm{V}$ from the fit with $c_4$ free, marginalized over the intrinsic-slope grid.
Blue dashed and dotted lines mark the $1\sigma$ and $3\sigma$ upper limits obtained with the physical prior $c_4\ge0$; the grey band indicates the region excluded at $3\sigma$.
Vertical coloured lines show the Milky Way and LMC values from the Drude fits of \citet{Li2008ApJ...685.1046L}; the SMC is bumpless in the same formalism and so coincides with zero.
\textit{Right:} the observed extinction curve in the vicinity of the bump (red points, for $\alpha_\lambda=1.5$) compared with the best-fit bump-free model (black), the same model with a bump at our $3\sigma$ upper limit (blue dashed), and with a Milky-Way-strength bump (purple dot-dashed).
The grey band marks the rest-frame position of the feature.
A Milky-Way-like bump is clearly inconsistent with the data.}}
\label{fig:bumplimit}
\end{figure*}

{The definition of $A_\mathrm{bump}/A_\mathrm{V}$, the value measured for \target, the upper limits and the reference values of \citet{Li2008ApJ...685.1046L} are given in Section~\ref{sec:result}; the full posterior is shown in the left panel of Fig.~\ref{fig:bumplimit}, where the small negative median is apparent.}
{We note that those formal limits are tighter than the detection threshold derived below, because $c_4$ fluctuated $0.8\sigma$ below zero; the conservative reading, which we adopt throughout, is that a Milky-Way-strength bump is firmly excluded while a feature as weak as the LMC bump would not have been detectable.}

{To answer the complementary question of what bump strength would have been detectable with these data, we performed an injection--recovery experiment.}
{Drude bumps of known amplitude were added to the \textit{observed} $A_\lambda$ data, thereby preserving the real, correlated noise and template-mismatch residuals, and the resulting data sets were refitted with the identical procedure, including the marginalization over $\alpha_\lambda$.}
{The recovery is linear and unbiased, with the recovered amplitude tracking the injected one offset by the $c_4=-0.0075$ baseline of the real data: injected bumps of $A_\mathrm{bump}/A_\mathrm{V}=0.10$, 0.20, 0.39, 0.59 and 1.06 (the last approximately the Galactic value) are recovered as $-0.04\pm0.17$, $+0.06\pm0.15$, $+0.26\pm0.14$, $+0.46\pm0.13$ and $+0.94\pm0.19$ respectively, i.e.\ at $-0.3\sigma$, $+0.4\sigma$, $+1.9\sigma$, $+3.4\sigma$ and $+5.1\sigma$ above zero (all values marginalized over the $\alpha_\lambda$ grid and computed from the unrounded posterior).}
{The $3\sigma$ detection threshold is $c_4\approx0.027$, i.e.\ $A_\mathrm{bump}/A_\mathrm{V}\approx0.5$, about half the Milky Way value, and a Milky-Way-strength bump would have been recovered at $5.1\sigma$ after marginalizing over $\alpha_\lambda$ (and at $8.0\sigma$ at fixed $\alpha_\lambda$).}
{This is illustrated in the right panel of Fig.~\ref{fig:bumplimit}.}

{We stress that the sensitivity is not limited by the modest resolving power of the PRISM.}
{At $z=4.556$ the bump falls at an observed wavelength of $1.21~\mu$m, where the reduced spectrum is sampled at $0.0173~\mu$m~pixel$^{-1}$ ($R\approx35$), whereas the Drude profile has an intrinsic rest-frame FWHM of $\approx486$~\AA\ ($\Delta\lambda/\lambda=0.224$, equivalent to $R\approx4.5$).}
{The feature would therefore span $\approx15$ spectral pixels, or roughly seven resolution elements, and is comfortably resolved.}
{The limiting factors are instead the covariance between $c_4$ and the smooth far-UV terms $(c_1,c_2,c_3)$, which can partially absorb a broad feature, and the marginalization over $\alpha_\lambda$; both are fully propagated into the limits quoted above.}
{Finally, we note that over the 1900--2500~\AA\ window relevant to the bump the rms residual of the bump-free fit is 0.09~mag, comparable to the 0.1~mag uncertainty floor adopted in the fit, so that these limits are not optimistic.}

\subsection{{Tests of the foreground-screen assumption}}

{Equation~(\ref{equ:3}) treats the dust as a uniform foreground screen.}
{We tested this assumption in four ways.}

{First, we compared the reddening of the broad-line region with that of the continuum.}
{A geometry in which a compact obscurer covers the accretion disc but not the much larger broad-line region would leave the broad lines far less extinguished than the adjacent continuum, and would therefore inflate their observed equivalent widths.}
{We therefore compared the observed rest-frame equivalent widths of four unblended broad lines with those of the intrinsic \citet{Selsing2016A&A...585A..87S} composite, measured over identical windows with the same local power-law continuum.}
{Because a reddening common to a line and its adjacent continuum cancels in the equivalent width, the ratio measures the differential reddening, $\Delta A_\lambda\equiv A_\lambda^\mathrm{cont}-A_\lambda^\mathrm{BLR}=2.5\log_{10}(\mathrm{EW_{obs}/EW_{int}})$.}
{For C\,{\sc iv}~1549, C\,{\sc iii}]~1909, Mg\,{\sc ii}~2799 and H$\alpha$~6563 the best-fit continuum extinctions are $A_\lambda=5.69$, 4.43, 1.88 and 0.57~mag (transmissions 0.0053, 0.0170, 0.1765 and 0.5927), and the measured equivalent-width ratios are 0.82, 1.70, 1.62 and 1.20, corresponding to $\Delta A_\lambda=-0.22$, $+0.58$, $+0.52$ and $+0.20$~mag.}
{The line- and continuum-emitting regions therefore see the same dust column to within $|\Delta A_\lambda|\le0.6$~mag out of up to 5.7~mag, a residual comfortably accounted for by the intrinsic object-to-object scatter in QSO line equivalent widths.}
{We omit H$\beta$ from this comparison because at the PRISM resolution it is blended with Fe\,{\sc ii} and with [O\,{\sc iii}]$\lambda\lambda$4959,5007, which the composite contains but \target does not (Section~\ref{sec:obs}); measured over 4700--4930~\AA\ it gives a ratio of 0.47, i.e.\ a line weaker than the composite, which is the opposite sense to the differential-covering scenario being tested and so does not affect the conclusion.}

{Second, we note that partially covering and clumpy geometries \textit{saturate}, and therefore flatten rather than steepen the observed curve.}
{For a leaky screen, $f_\lambda^\mathrm{obs}=f_\lambda^\mathrm{int}[(1-f_\mathrm{c})+f_\mathrm{c}e^{-\tau_\lambda}]$, the effective extinction tends to $-2.5\log(1-f_\mathrm{c})$ as $\tau_\lambda\to\infty$.}
{For a clumpy screen with Poisson-distributed clouds \citep{Natta1984ApJ...287..228N}, $A_\mathrm{eff}\propto\mathcal{N}(1-e^{-\tau_\mathrm{c}})$, and the effective steepness is $S_\mathrm{eff}=(1-e^{-S\tau_{\mathrm{c},V}})/(1-e^{-\tau_{\mathrm{c},V}})\le S$, with equality only in the optically thin limit.}
{An intrinsic $S=8$ would be degraded to $S_\mathrm{eff}=5.8$ for a per-clump optical depth $\tau_{\mathrm{c},V}=0.1$, and to $S_\mathrm{eff}=2.5$ for $\tau_{\mathrm{c},V}=0.5$.}
{Our measured value is thus a strict lower limit on the steepness of the underlying grain extinction law under any such geometry.}
{We note that this derivation averages the transmission over a finite source area, so it applies to a partially resolved or non-uniformly covered emitting region rather than to a strict point source; for a true point source a clumpy screen simply reproduces the intrinsic law along whichever sightline is realized.}
{The same caveat cuts the other way for the bump: \citet{Natta1984ApJ...287..228N} also show that a clumpy screen smooths discrete features, so an intrinsic 2175~\AA\ bump would be weakened by the same geometry that degrades $S$.}
{Our bump limit is therefore conservative only under the near-uniform screen established by the first and third tests; those tests, and in particular the bound $1-f_\mathrm{c}<0.007$, are what license it.}

{Third, we refitted the data with the covering fraction $f_\mathrm{c}$ as an additional free parameter.}
{With our flexible parametric law, the data drive $f_\mathrm{c}\to1$: the uncovered fraction is $1-f_\mathrm{c}=0.003$, with $1-f_\mathrm{c}<0.007$ at $3\sigma$.}
{This small leaked component ($\Delta\chi^2=137$ for one extra parameter, reducing to $\Delta\chi^2\approx30$ after the error rescaling described below) is plausibly residual template mismatch, host starlight, or scattered light; including it makes the inferred \textit{intrinsic} extinction law steeper still, $S=11.2_{-2.0}^{+6.8}$, again in the conservative direction.}
{Under this model $A_\mathrm{V}$ is correspondingly less well constrained ($0.64_{-0.27}^{+0.16}$), which is one reason we retain the pure screen as the fiducial case.}
{Conversely, with a fixed empirical law plus a free covering fraction the fits fail decisively, returning $\chi^2=3442$ for the SMC Bar and $\chi^2=18012$ for the Milky Way, against $\chi^2=1514$ for our fiducial screen fit to the same 338 data points.}
{Even after rescaling the uncertainties so that the fiducial fit has $\chi^2/\mathrm{dof}=1$, the partially covering SMC model remains disfavoured by $\Delta\chi^2\approx420$.}
{Partial covering therefore cannot manufacture the observed steepness from a standard extinction law.}

{Fourth, we considered the one geometry that could in principle steepen the emergent curve: a wavelength-dependent covering factor.}
{Because the disc emitting radius scales as $R\propto\lambda^{4/3}$, with $R(5500~\mathrm{\AA})/R(1500~\mathrm{\AA})=5.6$, an occulter whose projected size falls between the two would cover the far-UV emitting region more completely than the optical one.}
{This requires structure on scales of $5\times10^{14}$--$3\times10^{15}$~cm located at $\gtrsim1.9\times10^{18}$~cm, i.e.\ a transverse scale $10^{-3}$--$10^{-4}$ of the distance to the black hole.}
{It would also decouple the broad-line region from the continuum reddening, which the first test excludes, and produce a saturating rather than a monotonically rising blue end, which the third test excludes.}
{Since such a configuration would be transient, repeat spectroscopy searching for the variability it predicts (Section~\ref{subsec:smallgrains}) offers a direct observational test.}


\bibliography{main}{}
\bibliographystyle{aasjournalv7}



\end{document}